\documentclass[12pt,a4paper]{JHEP3}
\preprint{Cavendish--HEP--10/17\\CERN-PH-TH/2010-236\\
DAMTP-2010-82\\MCnet/10/19}

\usepackage{cite}
\usepackage{amssymb}
\usepackage{amsmath}
\usepackage{epsfig}
\usepackage{subfigure}
\usepackage{color}
\let\normalcolor\relax
\usepackage{graphicx}
\usepackage{inputenc}
\usepackage{xspace}
\usepackage{slashed}
\inputencoding{latin1}
\usepackage{epstopdf}
\usepackage{axodraw4j}

\def\gev{~{\rm GeV}}

\def\lsim{\mathrel{\raise.3ex\hbox{$<$\kern-.75em\lower1ex\hbox{$\sim$}}}}
\def\gsim{\mathrel{\raise.3ex\hbox{$>$\kern-.75em\lower1ex\hbox{$\sim$}}}}
\def\ifmath#1{\relax\ifmmode #1\else $#1$\fi}

\def\ptmx{p_{\rm miss}^x\,}
\def\ptmy{p_{\rm miss}^y\,}
\def\pmiss{{\bf p}_{\rm miss}}

\newcommand{\beq}{\begin{equation}}
\newcommand{\eeq}{\end{equation}}
\def\beqn{\begin{eqnarray}}
\def\eeqn{\end{eqnarray}}

\newcommand{\bea}{\begin{eqnarray}}
\newcommand{\eea}{\end{eqnarray}}

\newcommand{\Herwigpp}{H\protect\scalebox{0.8}{ERWIG++}\xspace}
\newcommand{\Pythia}{P\protect\scalebox{0.8}{YTHIA}\xspace}

\title{Searching for third-generation composite leptoquarks at the LHC}

\author{Ben Gripaios$^1$, Andreas Papaefstathiou$^2$, Kazuki Sakurai$^{2,3}$, Bryan Webber$^{2}$\\
$^1$CERN PH-TH, Geneva 23, 1211 Switzerland \\
$^2$Cavendish Laboratory, University of Cambridge, Cambridge, UK\\
$^3$Department of Applied Mathematics and Theoretical Physics, University of Cambridge, UK\\
Email: \email{gripaios@cern.ch, andreas@hep.phy.cam.ac.uk, sakurai@hep.phy.cam.ac.uk, webber@hep.phy.cam.ac.uk}
}

\abstract{Fermion masses may arise via mixing of elementary fermions with composite fermions of a strong sector in scenarios of strongly-coupled electroweak symmetry breaking. The strong sector may contain leptoquark states with masses as light as several hundred GeV. In the present study we focus on the scalar modes of such leptoquarks since their bosonic couplings are determined completely and hence their production cross sections only depend on their masses. We study all the possible gauge-invariant non-derivative and single-derivative couplings of the scalar leptoquarks to the quarks and leptons, which turn out to be, predominantly, of the third generation. We examine their phenomenology and outline search strategies for their dominant decay modes at the LHC.}


\begin{document}
\section {Introduction}
Any model of electroweak symmetry breaking must explain not only the masses of gauge bosons, but also those of the Standard Model (SM) fermions. In models where electroweak symmetry breaking is driven by strong dynamics, it is usual to assume that fermion masses arise in much the same way as in the SM. That is, fermion masses arise via a Yukawa-type coupling, but in which the SM Higgs is replaced by some scalar operator of the strong sector, $\mathcal{O}$, carrying the same gauge quantum numbers as the Higgs. However, this assumption necessarily makes it rather difficult, if not impossible, to satisfy constraints coming from flavour-physics experiments, whilst simultaneously providing a natural explanation of the hierarchy between the weak and Planck scales. The tension arises in the following way: suppose that flavour physics arises at some scale $\Lambda_f$. In order to suppress dangerous flavour-changing processes coming from four-fermion operators, one needs $\Lambda_f \gtrsim 10^3$ TeV. The mass of the top quark arises from an operator of the form
$$\mathcal{L} \subset \frac{\mathcal{O}Q t^c}{\Lambda_f^{d-1}}, $$ where $d \geq 1$ is the dimension of the operator $\mathcal{O}$. To generate the large top mass with $\Lambda_f \gtrsim 10^3$ TeV requires $d \lesssim 1.2-1.3$ \cite{Luty:2004ye}. Now, for the hierarchy problem to be solved, the theory should not contain any relevant operators that can be added to the Lagrangian. In particular, the dimension of any gauge-singlet operator arising in the operator product expansion of $\mathcal{O}^\dagger \mathcal{O}$ should roughly exceed four. This is clearly impossible in theories, such as weakly-coupled theories or strongly-coupled theories of large-N type, where correlation functions of operator products approximately factorize. Moreover, there are strong indications that it is also impossible in any theory which possesses a large hierarchy of scales and is thus is approximately a conformal theory in between \cite{Rattazzi:2008pe,Rychkov:2009ij,Rattazzi:2010yc}.

An alternative, proposed long-ago \cite{Kaplan:1991dc}, is to declare that the observed fermion masses arise by mixing of elementary fermions with composite, fermionic resonances of the strong sector. The latter are, of course, sensitive to the electroweak-breaking dynamics of the strong sector and transmit it to the elementary fermions via the mixing. Such a mechanism gives an automatic suppression of flavour-changing processes since the light fermions (for which the constraints from flavour experiments are strongest) are those which are least mixed with the flavour-changing dynamics of the strong sector. Moreover, it offers the hope that the observed hierarchies of masses and mixings of SM fermions may be related to the electroweak hierarchy, via strongly-coupling effects.

If fermion masses do arise in this way, it follows that the strongly-coupled sector knows not only about the $SU(2)_L \times U(1)_Y$ of electroweak symmetry of the SM (as it must, since it breaks it to the electromagnetic $U(1)_{em}$), but also about the $SU(3)$ of the colour interactions \cite{Gripaios:2009dq}. More precisely, the strongly-coupled sector must, at the very least, contain colour-triplet fermionic resonances that can mix with elementary colour-triplets to make the observed quarks. 

It is reasonable to expect that the strongly-coupled sector will contain other coloured resonances. If it contains bosonic, coloured resonances, then, depending on the particular gauge charges (or other global charges) these may be able to couple to a lepton and a quark, thus playing the role of leptoquark states.\footnote{Di-quark states are also possible, but we do not address them here.} For example, in the original model of \cite{Kaplan:1991dc},
the fermionic resonances arose as technibaryons of a technicolour $SU(3)$ interaction and (though not mentioned in \cite{Kaplan:1991dc}) such a model would contain leptoquarks arising as techni-mesons.

Such resonances will typically lie around the TeV scale, but may even be substantially lighter if they arise in the theory as pseudo-Nambu Goldstone bosons. Typically, such light states would be ruled out by flavour experiments, but in these models, leptoquark mediated flavour-changing processes are suppressed in exactly the same way as other flavour-changing processes that occur in the theory. Indeed, estimates given in~\cite{Gripaios:2009dq} suggest that such states may, as a result of this flavour suppression, be as light as the current Tevatron bounds of roughly 200 GeV. 

These composite leptoquarks therefore make an ideal target for LHC searches, and if discovered would give us a strong hint about the mechanism of electroweak symmetry breaking. 

Since the leptoquark couplings to light fermions are highly suppressed, the only relevant couplings for direct collider production and detection are those to third generation fermions.\footnote{For an alternative scenario with leptoquarks of this type, see \cite{Davidson:2010uu}.} As a result, the leptoquark states will  decay exclusively to third-generation fermions, that is to $t \tau$ or $t \nu_\tau$ or $b \tau$ or $b \nu_\tau$. Na\"{\i}vely, since the leptoquark couplings scale roughly with the Yukawa couplings, and since the Tevatron bounds preclude a leptoquark mass below $m_t$,\footnote{Searches at D0 for third generation scalar leptoquarks decaying exclusively to $b \tau$ or $b\nu_{\tau}$ yield bounds of 210 GeV \cite{Abazov:2008jp} and 229 GeV  \cite{Abazov:2007bsa} respectively.} one might conclude that decays involving the top must dominate. However, we shall see later that the gauge quantum numbers sometimes preclude couplings to top quarks and, of course, unknown global symmetries may also preclude one or more couplings. Thus we consider all four possible couplings. 

Since leptoquarks couple dominantly to third generation quarks and leptons, pair-production through colour gauge interactions will overwhelmingly dominate single production at the LHC. The channels of interest therefore involve pair-wise combinations of $t \tau$ or $t \nu_\tau$ or $b \tau$ or $b \nu_\tau$.\footnote{Note that third-generation lepton-quark couplings are also possible in $R$-parity-violating supersymmetric theories.}
The $2b 2\tau$ and $2b + \slashed{E_T}$ channels already have been the subject of searches at the Tevatron \cite{Abazov:2007bsa,Abazov:2008jp}, and can be adapted easily for the LHC. The use of novel kinematic variables such as $M_{T2}$ in this $2b+\slashed{E_T}$ channel may well improve the prospects for discovery and mass measurement. The two channels involving the top require more ingenuity, but merit investigation.

In the present paper we perform the first detailed phenomenological study of the possible production of such states at the LHC.  In Section~\ref{sec:phen} we briefly review their quantum numbers, couplings and decay modes, which we have implemented in the general-purpose event generator \Herwigpp~\cite{Bahr:2008pv,Bahr:2008tf}.\footnote{\Pythia~\cite{Sjostrand:2006za,Sjostrand:2007gs,Sjostrand:2008vc} contains an implementation of a single scalar leptoquark of arbitrary flavour.}  This allows us to propose and investigate some strategies for reconstructing third-generation leptoquark masses from their decay products, including those that involve top quarks, in Section~\ref{sec:strat}.  Our conclusions are presented in Section~\ref{sec:conc}.

Although we focus here on direct searches at the LHC, there are also promising channels for indirect searches, namely in 
$B_d \rightarrow K \bar{\mu} \mu$ and $B_s\rightarrow \mu \mu$ at LHCb, in $\mu \rightarrow e \gamma$ and $\tau \rightarrow \mu \gamma$, in $\mu-e$ conversion in nuclei, and in  $\tau \rightarrow \eta \mu$ at future B factories \cite{Gripaios:2009dq}.

\section{Phenomenology}\label{sec:phen}
\subsection{Scalar leptoquark pair-production}
We focus on scalar leptoquarks in the present study since their bosonic couplings are determined completely by QCD and hence their production cross sections only depend on their masses. Moreover, the lightest (and most easily accessible) leptoquarks in these scenarios arise as scalar, pseudo-Nambu Goldstone bosons. The type of leptoquarks we are considering are predominantly pair-produced via gluon-gluon fusion or quark-antiquark annihilation, due to the fact that they couple to the third generation quarks and leptons. Only charge-conjugate leptoquarks can be produced in this way: associated production of different leptoquarks is forbidden since it would not conserve the Standard Model gauge quantum numbers. Single production in association with a lepton is allowed but at a 14 TeV LHC it becomes dominant at leptoquark masses of about $2.2~\mathrm{TeV}$, at which point the total cross section is $\sigma \sim 10^{-2}~\mathrm{fb}$, already too low for discovery.
\subsubsection{Effective Lagrangian for interactions with gluons}
The effective Lagrangian describing the interaction of the scalar leptoquarks with gluons is~\cite{Blumlein:1996qp}
\begin{equation}
\mathcal{L}^g_S = \left(D^\mu_{ij} \Phi^j\right)^\dagger (D^{ik}_\mu \Phi_k)- M_{LQ}^2 \Phi^{i\dagger} \Phi_i\;\;,
\end{equation}
where $\Phi$ is a scalar leptoquark, $i,j,k$ are colour indices, the field strength tensor of the gluon field is given by
\begin{equation}
\mathcal{G}^a_{\mu \nu} = \partial_\mu \mathcal{A}_\nu^a - \partial_\nu \mathcal{A}^a_\mu + g_s f^{abc} \mathcal{A}_{\mu b} \mathcal{A}_{\nu c}\;\;,
\end{equation}
and the covariant derivative is
\begin{equation}
D^{ij}_\mu = \partial_\mu \delta^{ij} - i g_s t^{ij}_a \mathcal{A}_\mu^a\;\;.
\end{equation}
The Feynman rules that result from this Lagrangian and the
diagrams that contribute to pair-production of scalar leptoquarks are
given in Appendix~\ref{app:feynman}. Expressions for the cross
sections are given in Appendix~\ref{app:xsections}.
\subsection{Leptoquark decays}
\subsubsection{Non-derivative fermion couplings}\label{sec:naming}
The effective Lagrangian that describes the possible non-derivative couplings of the scalar leptoquarks to third-generation quarks and leptons is given by~\cite{Belyaev:2005ew}:
\begin{eqnarray}
\mathcal{L}_{nd} &=& ( g_{0L} \bar{q}_L^c i \tau _2 \ell_L + g_{0R} \bar{t}^c_R \tau_R ) S_0 \nonumber\\
&+& \tilde{g}_{0R} \bar{b}_R^c \tau_R \tilde{S}_0 + g_{1L} \bar{q}^c_L i \tau _2 \tau_a \ell _L S_1^a \nonumber \\
&+& (h_{1L} \bar{u}_R \ell_L + h_{1R} \bar{q}_L i\tau_2 \tau_R ) S_{1/2} + h_{2L} \bar{d}_R \ell_L \tilde{S}_{1/2} + \mathrm{h.c.}\;\;,
\label{eq:ells}
\end{eqnarray}
where the $\tau_a$ are the Pauli matrices, $q_L$ and $\ell_L$ are $SU(2)_L$ quark and lepton doublets respectively and $t_R$, $b_R$ and $\tau_R$ are the corresponding singlet fields. We denote charge conjugate fields by $f^c_{R,L} = (P_{R,L} f ) ^c$, where the superscript $^c$ implies charge conjugation. In Table~\ref{tb:lquarks} we give the quantum numbers for the five types of non-derivatively coupled scalar leptoquarks: the $SU(2)_L$-singlet complex scalars $S_0$, $\tilde{S}_0$, the $SU(2)_L$-triplet complex scalar $S_1$ and the $SU(2)_L$-doublets $S_{1/2}$ and $\tilde{S}_{1/2}$. 
\begin{table}[ptb]
\begin{center}
\begin{tabular}
{|c|c|c|c|c|c|c|} \hline
Name & $SU(3)_c$ & $T^3$ & $Y$  & $Q_{\mathrm{em}}$ & Decay mode & \Herwigpp id \\ \hline\hline
$S_0$ & $\bar{3}$ & 0 &  1/3 &   1/3 & $\bar{\tau}_R \bar{t}_R$, $\bar{\tau}_L \bar{t}_L$, $\bar{\nu} _{\tau ,L} \bar{b}_L$  & -9911561  \\ \hline\hline
$\tilde{S}_0$ & $\bar{3}$ & 0 & 4/3 &  4/3 &$ \bar{\tau}_R \bar{b}_R $ & -9921551  \\ \hline \hline
$S_1^{(+)}$ & $\bar{3}$ & +1 &  1/3 &  4/3 &$\bar{\tau}_L \bar{b}_L$  & -9931551  \\ \hline
$S_1^{(0)}$ & $\bar{3}$ & 0 & 1/3 &  1/3 & $\bar{\tau}_L\bar{t_L}$, $\bar{\nu}_{\tau,L} \bar{b}_L$ & -9931561  \\ \hline
$S_1^{(-)}$ & $\bar{3}$ & -1 &  1/3 & -2/3 &$\bar{\nu}_{\tau,L} \bar{t}_L$  & -9931661  \\ \hline\hline
$S_{1/2}^{(+)}$ & $3$ & +1/2 & 7/6 &  5/3 & $t_R \bar{\tau}_L, t_L \bar{\tau}_R$& 9941561  \\ \hline
$S_{1/2}^{(-)}$ & $3$ & -1/2 & 7/6 &  2/3 & $b_L \bar{\tau}_R$, $t_R \bar{\nu}_{\tau,L}$ & 9941551  \\ \hline\hline
$\tilde{S}_{1/2}^{(+)}$ & $3$ & +1/2 & 1/6 &  2/3 & $b_R \bar{\tau}_L$  & 9951551  \\ \hline
$\tilde{S}_{1/2}^{(-)}$ & $3$ & -1/2 & 1/6 &  -1/3 & $b_R \bar{\nu}_{\tau,L}$  & 9951651  \\ \hline
\end{tabular}
\end{center}
\caption{Numbering scheme, charges and possible decay modes for the non-derivatively coupled scalar leptoquarks. $Y$ represents the $U(1)_Y$ charge and $T^3$ is the third component of the $SU(2)_L$ charge. Since $S_1$ is an $SU(2)_L$ triplet, it contains three complex scalars. The $S_{1/2}$ and $\tilde{S}_{1/2}$ are $SU(2)_L$ doublets. The naming convention is explained in the text. The minus sign in the ids of some of the leptoquarks indicates the fact that they are anti-triplets of $SU(3)_c$.}
\label{tb:lquarks}
\end{table}

The numbering scheme used in our implementation of scalar leptoquarks in the \Herwigpp event generator is also given in Table~\ref{tb:lquarks}. The particles are numbered as $99NDDDJ$, where $N$ distinguishes the representation of the standard model gauge group, $DDD$ is the lowest possible number chosen to relate the leptoquark to the Particle Data Group (PDG) codes of decaying fermions, and $J = 2 S + 1$, where $S$ is the particle spin. The sign of the PDG code is negative for colour anti-triplets and positive for colour triplets. Hence $-9911561$ is the `first' type of leptoquark, $S_0$, and can decay to particles with codes $15$ ($\tau$) and $6$ ($t$). 

Notice that the first three kinds of leptoquarks, the $S_0$, $\tilde{S}_0$ and the $S_1$ triplet are colour anti-triplets and the particles (as opposed to the anti-particles) decay into an anti-lepton and an anti-quark. This is contrast to the $S_{1/2}$ and $\tilde{S}_{1/2}$ doublets, which are colour-triplets and decay into quarks and anti-leptons. 
\subsubsection{Derivative fermion couplings}\label{sec:derivcoup}
We also consider leptoquarks that couple derivatively to the quarks and leptons. The couplings of the leptoquarks to fermions involve three fields, and hence two independent positions for the derivative to act, modulo integration by parts. Here, we choose to put the derivative on either the quark or the lepton, such that the Lagrangian is given by:
\begin{eqnarray}
\mathcal{L}_d &=&  \frac{-i }{\sqrt{2} f} ( g'_{0L,i}\bar{q}_Lp^{\mu,i}  \gamma_\mu  \ell_L + g'_{0R,i} \bar{b}_Rp^{\mu,i}  \gamma_\mu \tau_R )S'_0 \nonumber\\ 
&+& \frac{-i }{\sqrt{2} f} \tilde{g}'_{0R,i} \bar{t}_R p^{\mu,i}  \gamma_\mu \tau_R \tilde{S}'_0 + \frac{-i }{\sqrt{2} f} g'_{1L,i} \bar{q}_Lp^{\mu,i}  \gamma_\mu  \tau _a \ell _L  S_1'^a \nonumber \\
&+& \frac{-i }{\sqrt{2} f} (h'_{1L,i} \bar{b}^c_R p^{\mu,i} \gamma_\mu \ell_L + h'_{1R,i} \bar{q}^c_L p^{\mu,i} \gamma_\mu \tau_R ) S'_{1/2} +\frac{-i }{\sqrt{2} f} h'_{2L,i} \bar{t}^c_R p^{\mu,i} \gamma_\mu \ell_L \tilde{S}'_{1/2} +\mathrm{h.c.}\;\;,\nonumber\\
\label{eq:lagd}
\end{eqnarray}
where the index $a \in \{1, 2, 3\}$ and $p^{\mu, i}$, $i \in \{l,q\}$, denotes the momentum of the lepton or quark.
\begin{table}[ptb]
\begin{center}
\begin{tabular}
{|c|c|c|c|c|c|c|} \hline 
Name & $SU(3)_c$ & $T^3$ & $Y$  & $Q_{\mathrm{em}}$ & Decay mode & \Herwigpp id \\ \hline\hline
$S_0'$ & $3$ & 0 &  2/3 & 2/3 & $t_R\bar{\nu}_{\tau,L}, b_R \bar{\tau}_L, b_L \bar{\tau}_R$ & 9961551  \\ \hline\hline
$\tilde{S}_0'$ & $3$ & 0 & 5/3 &  5/3 & $t_R \bar{\tau}_L, t_L \bar{\tau}_R$ & 9971561  \\ \hline \hline
$S_1^{'(+)}$ & $3$ & +1 &  2/3 & 5/3  & $t_R \bar{\tau}_L, t_L \bar{\tau}_R$  & 9981561  \\ \hline
$S_1^{'(0)}$ & $3$ & 0 & 2/3 & 2/3  & $t_R\bar{\nu}_{\tau,L}, b_L \bar{\tau}_R, b_R \bar{\tau}_L$ & 9981551  \\ \hline
$S_1^{'(-)}$ & $3$ & -1 &  2/3 & -1/3 & $b_R \bar{\nu}_L$ & 9981651  \\ \hline\hline
$S_{1/2}^{'(+)}$ & $\bar{3}$ & +1/2 & 5/6 & 4/3  & $\bar{b}_L \bar{\tau}_L, \bar{b}_R \bar{\tau}_R$  &-9991551  \\ \hline
$S_{1/2}^{'(-)}$ & $\bar{3}$ & -1/2 & 5/6 & 1/3  & $\bar{b}_L \bar{\nu}_{\tau,L}, \bar{t}_R \bar{\tau}_R, \bar{t}_L \bar{\tau}_L$  &-9991561    \\ \hline\hline
$\tilde{S}_{1/2}^{'(+)}$ & $\bar{3}$ & +1/2 & -1/6 & 1/3  &$\bar{t}_L \bar{\tau}_L, \bar{t}_R \bar{\tau}_R$  & -9901561   \\ \hline
$\tilde{S}_{1/2}^{'(-)}$ & $\bar{3}$ & -1/2 & -1/6 & -2/3  & $\bar{t}_L \bar{\nu}_{\tau,L}$  & -9901661  \\ \hline
\end{tabular}
\end{center}
\caption[]{Numbering scheme, charges and possible decay modes for the derivatively-coupled scalar leptoquarks. The details are as in Table~\ref{tb:lquarks}.}
\label{tb:lquarksprime}
\end{table}

The charges of the primed scalar states appear in Table~\ref{tb:lquarksprime}; they correspond, of course, to those of vector leptoquarks. Notice that whereas the $S_0$ is a colour anti-triplet, $S'_0$ is a colour triplet and so on.

Consider a leptoquark $S_0'$ that couples derivatively to fermions in the following way:
\begin{equation}
\mathcal{L} \sim \frac{1 }{\sqrt{2} f} \left( g'_{0L,i} \bar{t}_L \slashed{p}^i S_0' \nu_L + g'_{0L,i} \bar{b}_L \slashed{p}^i S_0' \tau_L +g'_{0R,i}\bar{b}_R \slashed{p}^iS_0' \tau_R \right)+\mathrm{h.c.}\;\;, 
\end{equation}
where the $f$ is the sigma-model scale for the strong dynamics. Consider the decay of the $S_0'$ to on-shell fermions via the coupling $g'_{0L,i}\bar{b}_L p^{\mu,i}  \gamma_\mu  t_L$. We then have:
\begin{eqnarray}
g'_{0L,i} \bar{b}_L \slashed{p}^i \tau_L S_0 '  &=& g'_{0L,q} \bar{b}_L \slashed{p}^q S_0' \ell_L + g'_{0L,\ell}\bar{b}_L \slashed{p}^\ell \tau _L S_0' \nonumber\\
&=& g'_{0L,q} m_b \bar{b}_R S_0 ' \ell_L + g'_{0L,\ell} m_\tau \bar{b}_L S_0' \tau_R\;\;.
 \label{eq:noconjmanip}
\end{eqnarray}
Note that the chirality of one decay product is reversed in each term by the mass insertion, which breaks the gauge symmetry. An equivalent manipulation is given in Appendix~\ref{app:conj} for terms that contain conjugate fields. For simplicity of the analysis, we choose to set the quark and lepton primed couplings for each term equal, $g'_\ell = g'_q = g'$, where $g'$ represents $g'_0$, $g'_1$ or $h'_1$, for the rest of the paper. 
As a result of the above manipulation, an effective Lagrangian for the on-shell decay of a scalar leptoquark $S'_0$ may be written as:
\begin{eqnarray}
\mathcal{L}_{eff.} &\sim& \frac{1 }{\sqrt{2} f} \left( g'_{0L} m_t \bar{t}_R S_0' \nu_{\tau,L} + g'_{0L} m_b \bar{b}_R S_0' \tau_L + g'_{0L} m_\tau \bar{b}_L S_0' \tau_R \right. \nonumber\\ 
&+&\left.g'_{0R}\bar{b}_L m_b S_0' \tau_R+ g'_{0R}\bar{b}_R m_\tau S_0' \tau_L \right) +\mathrm{h.c.} \\
\Rightarrow \mathcal{L}_{eff.} &\sim& \left[ \frac{1 }{\sqrt{2} f}  (g'_{0L} m_b+ g'_{0R} m_\tau)\right] \bar{b}_R S_0' \tau_L\nonumber\\
&+& \left[ \frac{1 }{\sqrt{2} f} ( g'_{0L} m_\tau + g'_{0R} m_b )  \right] \bar{b}_L S_0' \tau_R \nonumber \\
&+& \left[  \frac{1 }{\sqrt{2} f} (g'_{0L} m_t)\right]  \bar{t}_R S_0' \nu_{\tau,L}+\mathrm{h.c.}\;\;,
\end{eqnarray}
effectively converting all the derivative couplings to ones that look like those for the unprimed leptoquarks, with the lepton or fermion masses appearing in the coupling. See Appendix~\ref{app:conj} for the full effective Lagrangian. Since the scale $f$ is typically a few hundred GeV, couplings proportional to the top quark mass are expected to dominate when the corresponding decays are kinematically allowed. The on-shell fermion assumption is realistic since the widths of the fermions are small in comparison to their masses and hence off-shell effects are negligible.
\subsubsection{Decay widths}
The decay width of non-derivatively coupled scalar leptoquarks in the limit of \textit{massless} quarks and leptons can be calculated by~\cite{Belyaev:2005ew}: 
\begin{equation}
\Gamma = \frac{M_{LQ}}{16 \pi} \left( \lambda^2_L (\ell q) + \lambda^2_L (\nu q) + \lambda_R^2 (\nu q) \right) \;\;,
\label{eq:width}
\end{equation}
where the couplings $\lambda_{L,R}(\ell q)$ for the types of leptoquarks we are considering are given in Table~\ref{tb:lambdas}  in terms of the couplings that appear in the Lagrangian. The couplings are taken to be real. The expression gives, for quark-lepton couplings $g \sim 0.1$ and leptoquark mass of $\sim 400 \gev$, a width of $\sim 0.1 \gev$.
\begin{table}[ptb]
\begin{center}
\begin{tabular}
{|c|c|c|c|} \hline
Name & $\lambda_{L} (\ell q)$ & $\lambda_R(\ell q)$ & $\lambda_L(\nu q) $    \\ \hline\hline
$S_0$ & $g_{0L}$ & $g_{0R}$ & $-g_{0L}$ \\ \hline\hline
$\tilde{S}_0$ & 0 & $\tilde{g}_{0R}$ & 0 \\ \hline \hline
$S_1^{(+)}$ & $\sqrt{2} g_{1L}$ & 0  &  0 \\ \hline 
$S_1^{(0)}$ & $-g_{1L} $ & 0 &  $-g_{1L}$ \\ \hline
$S_1^{(-)}$ & 0 & 0  & $\sqrt{2} g_{1L}$ \\ \hline \hline
$S_{1/2}^{(+)}$ & $h_{1L}$ & $h_{1R}$ & 0 \\ \hline
$S_{1/2}^{(-)}$ & 0 & $-h_{1R}$ & $h_{1L}$ \\ \hline \hline
$\tilde{S}_{1/2}^{(+)}$ & $h_{2L}$ & 0 & 0 \\ \hline 
$\tilde{S}_{1/2}^{(-)}$ & 0& 0 & $h_{2L}$ \\ \hline
\end{tabular}
\end{center}
\caption{The $\lambda_i$ couplings of the non-derivatively scalar leptoquarks to the different quark-lepton combinations, as they appear in the Lagrangian.}
\label{tb:lambdas}
\end{table}
The decay width to massive $q \ell$ is further suppressed by a phase space factor compared to the massless quark and lepton width~\cite{Abazov:2007bsa}:
\begin{eqnarray}
F \sim ( 1 - r_q - r_\ell )\sqrt{ 1+(r_q-r_\ell)^2 - 2 r_q - 2 r_l } \;\;,
\end{eqnarray} 
where $r_{q,\ell}$ are the squared ratios $ m_{q,\ell}^2/M_{LQ}^2$ respectively.

\begin{table}[htb]
\begin{center}
\begin{tabular}
{|c|c|c|c|} \hline
Name & $\lambda_{L} (\ell q) \times \sqrt{2} f$ & $\lambda_R(\ell q)  \times \sqrt{2} f$ & $\lambda_L(\nu q)  \times \sqrt{2} f $    \\ \hline\hline
$S_0'$ & $ g'_{0L,q} m_b + g'_{0R,\ell} m_{\tau} $ & $g'_{0R,q} m_b + g'_{0L,\ell} m_{\tau} $ & $ g'_{0L,q} m_t $ \\ \hline\hline
$\tilde{S}_0'$ & $\tilde{g}'_{0R,\ell} m_{\tau}$ & $\tilde{g}'_{0R,q} m_t$ & 0 \\ \hline \hline
$S_1^{'(+)}$ & $\sqrt{2} g'_{1L,q} m_t $ & $\sqrt{2} g'_{1L,\ell} m_{\tau}$ &  0 \\ \hline 
$S_1^{'(0)}$ & $- g'_{1L,q} m_b $ & $-g'_{1L,\ell} m_\tau$ &  $g'_{1L,q} m_t $ \\ \hline
$S_1^{'(-)}$ & 0 & 0  & $\sqrt{2} g'_{1L,q} m_b$ \\ \hline \hline
$S_{1/2}^{'(+)}$ & $h'_{1L,q} m_b + h'_{1R,\ell} m_{\tau} $ & $h'_{1R,q}  m_b + h'_{1L,\ell} m_{\tau} $  & 0  \\ \hline
$S_{1/2}^{'(-)}$ & $h'_{1R,\ell} m_{\tau} $ & $h'_{1R,q} m_t$ & $h'_{1L,q} m_b$ \\ \hline \hline
$\tilde{S}_{1/2}^{'(+)}$  & $h'_{2L,\ell} m_{\tau}$ & $h'_{2L,q} m_t$  & 0 \\ \hline
$\tilde{S}_{1/2}^{'(-)}$ & 0 & 0 & $h'_{2L,\ell} m_t$ \\ \hline
\end{tabular}
\end{center}
\caption{The $\lambda_i$ couplings of the derivatively-coupled
  (primed) scalar leptoquarks to the different quark-lepton
  combinations, as they appear in the Lagrangian. In our analysis, we have set the
  quark and lepton couplings equal for simplicity.}
\label{tb:lambdasprime}
\end{table}
\begin{table}[htb]
\begin{center}
\begin{tabular}
{|c|c|c|} \hline
Decay mode & Decay width (GeV) & BR   \\ \hline\hline
$\bar{S}_0\rightarrow  \tau^- t$ & 0.1040 & 0.5666 \\ \hline
$\bar{S}_0 \rightarrow \nu_\tau b$ & 0.07956 & 0.4334  \\ \hline \hline
$\bar{\tilde{S}}_0 \rightarrow \tau^- b$ & 0.07956 & 1  \\ \hline \hline
$\bar{S}_1^{(+)} \rightarrow \tau^- b$ & 0.1591 & 1  \\ \hline 
$\bar{S}_1^{(0)} \rightarrow \tau^- t$ & 0.05225 & 0.3964  \\ \hline 
$\bar{S}_1^{(0)} \rightarrow \nu_{\tau} b$ & 0.07956 & 0.6036  \\ \hline 
$\bar{S}_1^{(-)} \rightarrow  \nu_{\tau} t$ & 0.1045 & 1  \\ \hline \hline
$S_{1/2}^{(+)} \rightarrow  \tau^+ t$ & 0.1040 & 1  \\ \hline 
$S_{1/2}^{(-)} \rightarrow  \tau^+ b$ &  0.07956 & 0.6036  \\ \hline 
$S_{1/2}^{(-)} \rightarrow  \bar{\nu}_\tau t$ & 0.05225 & 0.3964  \\ \hline \hline
$\tilde{S}_{1/2}^{(+)} \rightarrow  \tau^+ b$ & 0.07956 & 1  \\ \hline 
$\tilde{S}_{1/2}^{(-)} \rightarrow  \bar{\nu}_\tau b$ & 0.07956 & 1  \\ \hline 
\end{tabular}
\end{center}
\caption{Decay widths for non-derivatively coupled scalar leptoquarks of mass $M_{LQ} = 400\gev$ and couplings $g = 0.1$.}
\label{tb:widths}
\end{table}
Table~\ref{tb:lambdasprime} shows the couplings for the primed, derivatively-coupled, scalar leptoquarks. The expression for the width given in Eq.~(\ref{eq:width}) remains unchanged in the case of the primed leptoquarks, with the couplings $\lambda_i$ taking the appropriate values.
Tables~\ref{tb:widths} and~\ref{tb:dwidths} show example decay widths and branching ratios for scalar leptoquarks of mass $M_{LQ} = 400 \gev$ and couplings $g = 0.1$. In the case of derivatively coupled leptoquarks we choose a suppression scale $f = 800 \gev$.

\begin{table}[ptb]
\begin{center}
\begin{tabular}
{|c|c|c|} \hline
Decay mode & Decay width (GeV) & BR   \\ \hline\hline
$S'_0 \rightarrow  \tau^- b$ & $4.440\times 10^{-6}$ & 0.0036 \\ \hline 
$S'_0 \rightarrow \nu_\tau t$ & 0.001239 & 0.9964  \\ \hline \hline
$\tilde{S}'_0 \rightarrow \tau^- t$ & 0.001239 & 1  \\ \hline \hline
$S_1'^{(+)} \rightarrow \tau^- t$ & 0.002478 & 1  \\ \hline 
$S_1'^{(0)} \rightarrow \tau^- b$ & $1.292\times 10^{-6}$ & 0.0010  \\ \hline 
$S_1'^{(0)} \rightarrow \nu_{\tau} t$ & 0.001239 & 0.9990  \\ \hline 
$S_1'^{(-)} \rightarrow  \nu_{\tau} b$ & $2.193\times 10^{-6}$ & 1  \\ \hline \hline
$\bar{S}_{1/2}'^{(+)} \rightarrow  \tau^- b$ & $4.440\times10^{-6}$ & 1  \\ \hline 
$\bar{S}_{1/2}'^{(-)} \rightarrow  \tau^- t$ & 0.001239 & 0.9991  \\ \hline 
$\bar{S}_{1/2}'^{(-)} \rightarrow  \nu_\tau b$ & $1.098\times 10^{-6}$ & 0.0009 \\ \hline \hline
$\bar{\tilde{S}}_{1/2}'^{(+)} \rightarrow  \tau^- t$ & 0.001234 & 1  \\ \hline 
$\bar{\tilde{S}}_{1/2}'^{(-)} \rightarrow  \nu_\tau t$ & 0.001239 & 1  \\ \hline 
\end{tabular}
\end{center}
\caption{Decay widths for derivatively-coupled (primed) scalar leptoquarks of mass $M_{LQ} = 400\gev$, couplings $g' = 0.1$ and suppression scale $f = 800 \gev$.}
\label{tb:dwidths}
\end{table}

\section{Experimental search strategies}\label{sec:strat}
\subsection{Reconstruction strategies}\label{sec:outlinestrategy}
Table~\ref{tb:strategy} provides an overview of our suggested reconstruction strategies for the different types of leptoquarks. The `stransverse' mass variable which appears in the table, $M_{T2}$, has been defined previously in~\cite{Lester:1999tx}, for the case of identical semi-invisible pair decays as:
\begin{equation}
M_{T2} \equiv
     \min_{ \slashed{\bf c}_T + \slashed{\bf c}'_T = \slashed{\bf p_T} }
     \left\{
     \max { \left( M_T, M_T' \right) }
       \right\} \;\;,
\end{equation}\label{eq:mt2}
where the minimisation is taken over $\slashed{\bf c}_T$ and $\slashed{\bf c}'_T$, the transverse momenta of the invisible particles, with the constraint that their sum equals $\slashed{\bf p_T}$, the total missing transverse momentum, and $M_T$ and $M_T'$ are the transverse masses calculated for the two decay chains. We assume that the invisible particles are massless and use the jet masses in our definitions of $M_{T2}$. The new variables $M_{\mathrm{min}}^{\mathrm{bal}}$ and $M_{\mathrm{min}}$ will be defined in section~\ref{sec:ttaubnurecon}.

We present our analysis of the mass reconstruction techniques for each pair-production decay mode separately, initially at parton level and then at detector level, including discussion of the relevant backgrounds. We focus on the $S_0$ singlet, $S_1$ triplet and $S_{1/2}$ doublet and outline how to generalise the strategy to all the leptoquark multiplets.

It is evident from Tables~\ref{tb:widths} and~\ref{tb:dwidths} that the leptoquark decay widths are generally much smaller than the resolution of the detector components, and hence our analysis is not sensitive to the decay widths. Throughout what follows we have set the leptoquark couplings to fermions to the value  $g=0.1$. This value represents an estimate of the leptoquark couplings to third generation quarks and leptons, derived using the measured fermion masses as explained in~\cite{Gripaios:2009dq}. The resulting width-to-mass ratio for the leptoquarks corresponding to this coupling, according to Eq.~(\ref{eq:width}), is $\mathcal{O}(10^{-4})$.

We use the \Herwigpp implementation of the model to generate a number of events corresponding to an integrated luminosity of $10$ fb$^{-1}$ of the relevant signal and $t\bar{t}$ background samples. Subsequently we use the \texttt{Delphes} framework~\cite{Ovyn:2009tx} to simulate the detector effects and assess the
feasibility of reconstruction in an experimental
situation.\footnote{\texttt{Delphes} is a framework for fast
  simulation of a general-purpose collider experiment.}
\texttt{Delphes} includes the most crucial experimental features:
geometry of the central detector, the effect of the magnetic field on the
tracks, reconstruction of photons, leptons, $b$-jets, $\tau$-jets and
missing transverse energy. It contains simplifications such as
idealised geometry, no cracks and no dead material. We use the default
parameter settings in the \texttt{Delphes} package that correspond to
the ATLAS detector. Critical features of our analysis are
both $b$- and $\tau$-tagging of jets and we caution the reader to take into
consideration that the relevant efficiencies contain a degree of
uncertainty at this stage of the LHC experiment and for the near
future. The $b$-tagging present in the \texttt{Delphes} framework assumes an
efficiency of 40\% if the jet has a parent $b$-quark, 10\% if the jet
has a parent $c$-quark and 1\% if the jet is light (i.e. originating
from $u$, $d$, $s$ or $g$). The identification of hadronic $\tau$-jets is consistent
with the one applied in a full detector simulation. The resulting efficiencies for
hadronic $\tau$-jets are in satisfactory agreement with those
assumed by ATLAS and CMS. See~\cite{Ovyn:2009tx} for further details.

Throughout the analysis we
apply transverse momentum cuts of at least $30 \gev$. Since we are always working with high-transverse momentum objects,
we can assume that pile-up arising due to secondary proton-proton
collisions is under experimental control. See, for example, the ATLAS $t\bar{t} H (\rightarrow b\bar{b})$ study in~\cite{Aad:2009wy}.
\begin{table}[ptb]
\begin{center}
\begin{tabular}
{|c|c|c|c|c|} \hline
modes & types & technique \\ \hline\hline 
$(t\tau)(b \nu)$ & $S_0$, $S_1^{(0)}$  & $j_{\tau} \parallel \nu_{\tau}$, mass constraints \\
      &       &  $\Rightarrow$ edge reconstruction ($M_{\mathrm{min}}^{\mathrm{bal}}$, $M_{\mathrm{min}}$, $M_{T2}$)       \\ \hline
$(t\tau) (t\tau)$& $S_0$, $S_1^{(0)}$,   & two $j_{\tau} \parallel \nu_{\tau}$, mass constraints  \\   
      &    $ S^{(+)}_{1/2}$, $\tilde{S}'_0$     &  $\Rightarrow$ full reconstruction      \\ \hline
 $(b \nu) (b \nu)$ & $S_0$, $S_1^{(0)}$,    & $M_{T2}$ \\
      &   $\tilde{S}^{(-)}_{1/2}$, $S_1'^{(-)}$    &            \\ \hline
$(b\tau) (b\tau)$ & $S^{(+)}_1$, $\tilde{S}_{1/2}^{(-)}$ & two $j_{\tau} \parallel \nu_{\tau}$, mass constraints   \\
              & $\tilde{S}_0$, $S_{1/2}'^{(+)}$, & $\Rightarrow$ full reconstruction    \\
              & $S_1'^{(0)}$ &  \\ \hline
$(t\nu) (t\nu)$ & $S^{(-)}_1$, $S_{1/2}^{(-)}$ &  \\
              & $S'_0$, $S_{1}'^{(0)}$, &$M_{T2}$    \\
              & $\tilde{S}_{1/2}'^{(-)}$ &  \\ \hline
$(t\nu) (b\tau)$ & $S^{(-)}_{1/2}$, $S_{0}'$ & $j_{\tau} \parallel \nu_{\tau}$, mass constraints   \\
              & $S_1'^{(0)}$ & $\Rightarrow$ edge reconstruction ($M_{\mathrm{min}}^{\mathrm{bal}}$, $M_{\mathrm{min}}$, $M_{T2}$)         \\ \hline
\end{tabular}
\end{center}
\caption{The table outlines the general reconstruction strategy for leptoquark
  pair-production for the different types of leptoquarks. For variable definitions and further details see the respective sections.}
\label{tb:strategy}
\end{table}

\subsection[$(t\tau)(t\tau)$ decay mode]{\boldmath $(t\tau)(t\tau)$ decay mode}\label{sec:tttt}
\begin{figure}[!t]
  \centering 
    \includegraphics[scale=0.60]{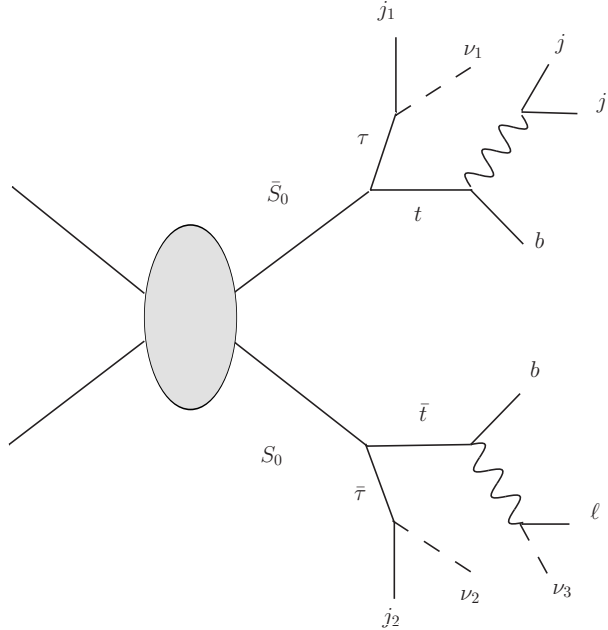}
\caption{Pair-production of $S_0$ leptoquarks with decay to $(t\tau)(t\tau)$, followed by one hadronic and one semi-leptonic top decay.} 
\label{fig:s0topology}
\end{figure}

\begin{figure}[!t]
  \centering 
  \vspace{1.5cm}  
  \hspace{4.0cm}
    \includegraphics[scale=0.60, angle=90]{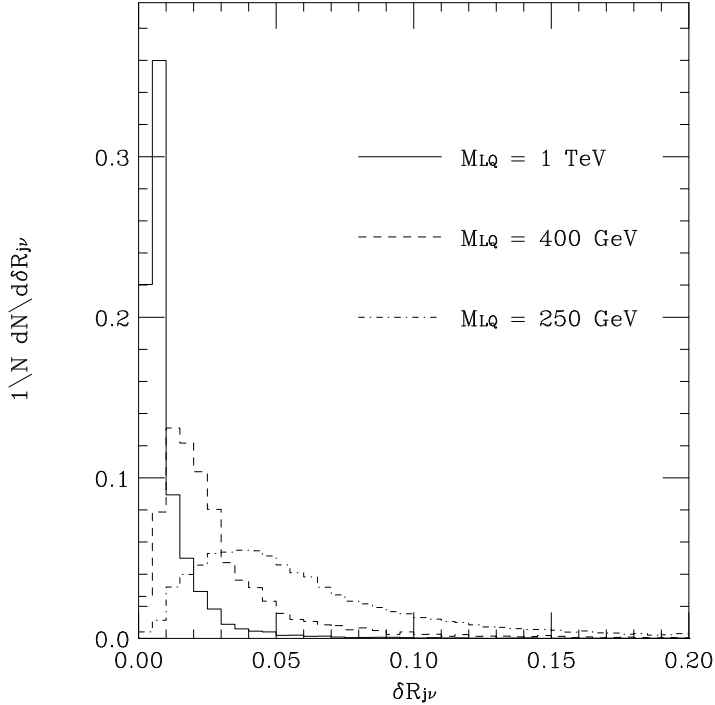}
  \vspace{0.5cm}
\caption{The distribution of the distance in $R$-space ($\delta R =
  \sqrt { \delta \eta ^2 + \delta \phi ^2 }$) between the momenta of
  the $\tau$ jet and the $\tau$ invisibles in $S_0$ pair-production
  for $M_{LQ} = 1, 0.4, 0.25$ TeV.}
\label{fig:tauangle}
\end{figure}

We examine the possibility of full reconstruction of the
topology shown in Figure~\ref{fig:s0topology}, where we have, for example. $S_0 (\bar{S}_0)
\rightarrow bjj j_1 \nu_1 $ and $S_0 (\bar{S}_0) \rightarrow b \ell \nu_3 j_2
\nu_2$, where $\nu_1$ and $\nu_2$ represent one or more neutrinos coming
from the $\tau$ decays and $\ell$ can be either a muon or an
electron. We can assume that the neutrinos $\nu_{1,2}$ associated with the decays
of the $\tau$s are collinear with the direction of the jets $j_{1,2}$ associated
with them. The validity of this assumption has been confirmed using \Herwigpp, for leptoquarks of mass 1, 0.4 and 0.25 TeV, as may be seen in Figure~\ref{fig:tauangle}, which shows the distribution of $\delta R =
\sqrt { \delta \eta ^2 + \delta \phi ^2 }$ between the momenta of the
$\tau$ jet partons and the $\tau$ invisibles. The assumption is employed in our reconstruction of any leptoquark decay mode containing a $\tau$-jet.

The top quark branching ratios are $\sim 0.216$ for the
semi-leptonic $e, \mu$ modes and $\sim 0.676$ for the hadronic top
modes, this is multiplied by two (since we have either the $t$ or
$\bar{t}$ in each of these) resulting in a $\sim 0.292$ factor just
for the top decay modes. The branching ratios and cross sections for
$S_0 \bar{S}_0$ production depend on the leptoquark mass and coupling and are shown in
Table~\ref{tb:s0br}, where the last column is the resulting cross
section for the whole process under study. We focus on $400\gev$ leptoquarks since these are clearly not excluded by direct searches and still provide a sufficient number of events to be potentially discovered at a reasonable luminosity (10 fb$^{-1}$) at $14$ TeV.
\begin{table}[!t]
\begin{center}
\begin{tabular}
{|c|c|c|c|} \hline
$M_{S_0}$ (GeV) & $\sigma(pp \rightarrow S_0\bar{S}_0)$ (pb) & BR($t\tau$) & $\sigma(t\tau \bar{t}\bar{\tau} \rightarrow b \bar{b} jj \ell(=e,\mu) \nu \tau \bar{\tau})$ (pb)  \\ \hline\hline 
174.2 (= $m_{top}$) & 141(1)   &   0.  & 0. \\ \hline
250 & 24.3(3) & 0.34 & 0.729 \\ \hline
400 & 2.000(7) & 0.567 & 0.188  \\ \hline
500 & 0.561(6) &  0.606 & 0.06  \\ \hline 
1000 &  5.94(7) $\times 10^{-3}$& 0.65 & $7.3\times 10^{-4}$  \\ \hline  \hline
$M_{top}$ (GeV) & $\sigma(pp \rightarrow t\bar{t})$ (pb) & - & $\sigma(t\bar{t} \rightarrow b \bar{b} jj \ell(=e,\mu) \nu)$ (pb) \\ \hline 
174.2 & 834(1) & - & 242 \\\hline 
\end{tabular}
\end{center}
\caption{$S_0\bar{S}_0$ total cross section at the LHC at 14 TeV pp c.o.m energy, branching ratio to
  $t\tau$ and remaining cross section taking into account the top branching ratios. The corresponding $t\bar{t}$ values are given for comparison.}
\label{tb:s0br}
\end{table}

\subsubsection{Kinematic reconstruction}\label{sec:ttauttaurecon}

The final states of $S_0 \bar S_0 \to \bar t \tau^+ t \tau^-$ processes contain 
many decay products including neutrinos.
If the system has a large enough number of kinematical constraints, 
such as mass-shell conditions and balance of 
the total transverse momentum, we can completely reconstruct the kinematics of the system.
The numbers of unknown variables and constraints are summarised in Table~\ref{tab:ttauttau} 
for each decay pattern of the tops:
(1) both tops decay hadronically, (2) one top decays semi-leptonically and another hadronically and (3)
both tops decay semi-leptonically.
\begin{table}[!t]
\begin{center}
\begin{tabular}{|l|c|c|}
\hline
Decay type&\# of unknowns&\# of constraints\\
\hline
(1) had,had&$1+(0+2)N$&$(2+2)N$\\
\hline
(2) had,lep &$1+ (4+2)N$&$(5+2)N$\\
\hline
(3) lep,lep&$1+(8+2)N$&$(8+2)N$\\
\hline
\end{tabular}
\caption{The numbers of unknown variables 
([$m_{LQ}$], [$\nu$ from top], [energy fraction of tau])
and constraints ([mass-shell conditions], [balance of missing momentum]) 
in $N$ events of each decay type.
The mass-shell conditions that could constrain the unknown variables
are counted here, i.e.~the mass-shell conditions on $S_0$, leptonic top, $W$ and 
$\nu$ from leptonic top decay. 
}
\label{tab:ttauttau}
\end{center}
\end{table}%
As mentioned above, we assume 
$\tau$ neutrinos are collinear to the $\tau$-jets,
leaving two unknown parameters associated with the taus, namely the
energy ratios $z_i$ ($i=1,2$, $z_i \ge 1$) which are defined (neglecting masses) by:
\begin{eqnarray}\label{eq:zis}
p_{\tau_i} &=& z_i p_{j_i}
\nonumber \\
p_{\nu_i} &=& p_{\tau_i} - p_{j_i} = (z_i - 1) p_{j_i} \;,
\end{eqnarray}
where $p_{\tau_i}$, $p_{j_i}$ and $p_{\nu_i}$ are the four-momenta of the $\tau$ leptons, $\tau$-jets and $\tau$ neutrinos,
respectively. 
With this assumption, the unknown variables 
in Table~\ref{tab:ttauttau} are the mass of the leptoquark, the four momenta of neutrinos
from leptonic top decays and the energy fractions associate with the neutrinos from the tau decays.
The mass-shell conditions that could constrain the unknown variables
are counted in Table~\ref{tab:ttauttau},
i.e.~the mass-shell conditions on $S_0$, leptonic top, $W$ and 
$\nu$ from leptonic top decay.

As can be seen, 
we can wholly reconstruct the kinematics of a single event
only in decay types (1) and (2). 
In decays of type (1), it would be difficult to reconstruct both hadronic tops 
because of the large combinatorial background. 
Thus, we focus on decay type (2) and attempt to determine the leptoquark mass.
As we show in Appendix~\ref{app:ttauttau}, in this case one obtains a quartic equation for the energy ratio $z_2$, and hence in general up to four solutions for the leptoquark mass, at least one of which should be close to the true value if the visible momenta and missing transverse momenta are well measured.

\subsubsection{Parton-level reconstruction}
\begin{figure}[!htb]
  \centering 
  \vspace{1.5cm}
  \hspace{3.0cm}
 \includegraphics[scale=0.35, angle=90]{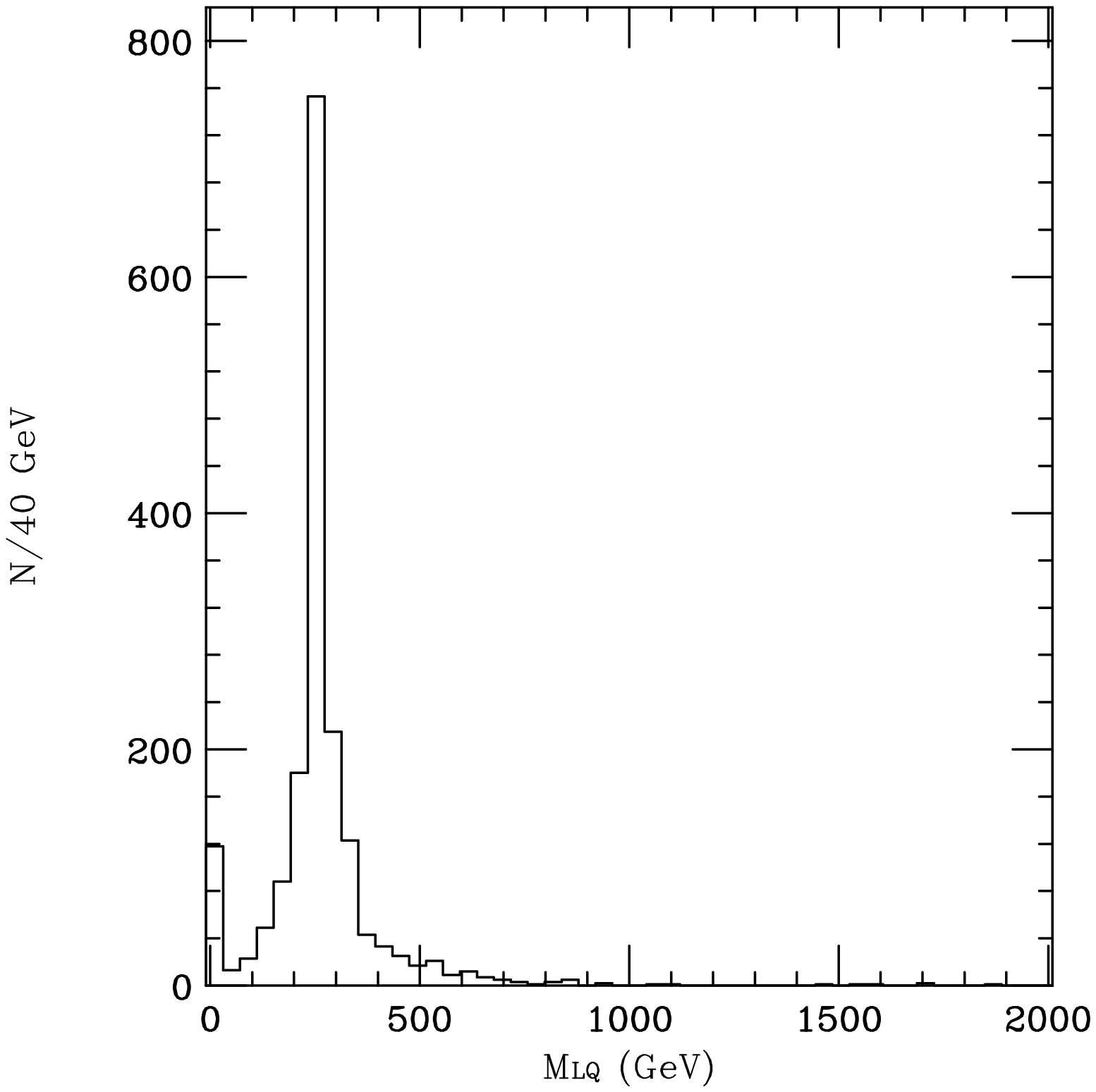}
  \hspace{3.5cm}
  \includegraphics[scale=0.35, angle=90]{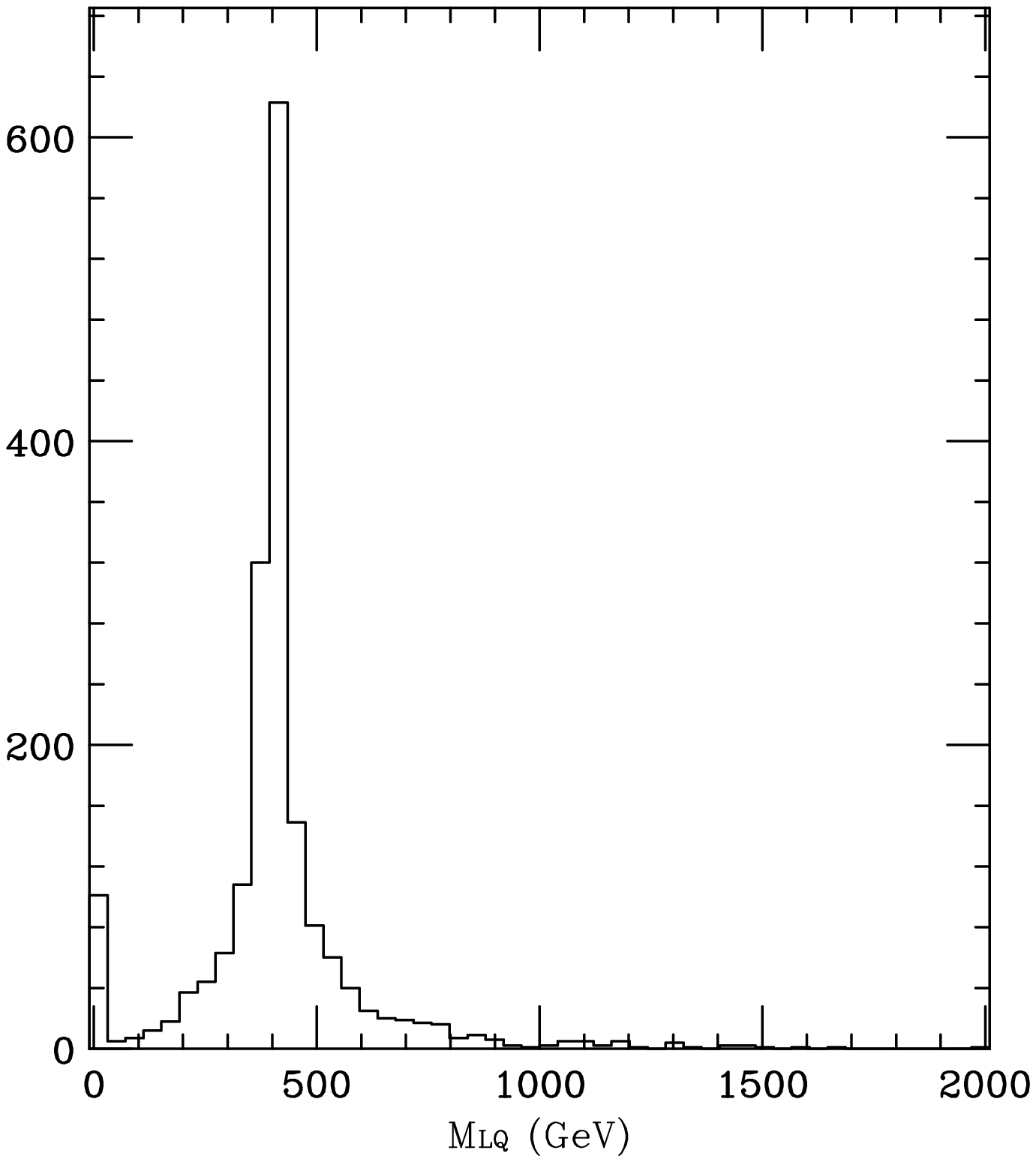}
  \hspace{3.5cm}
 \includegraphics[scale=0.35, angle=90]{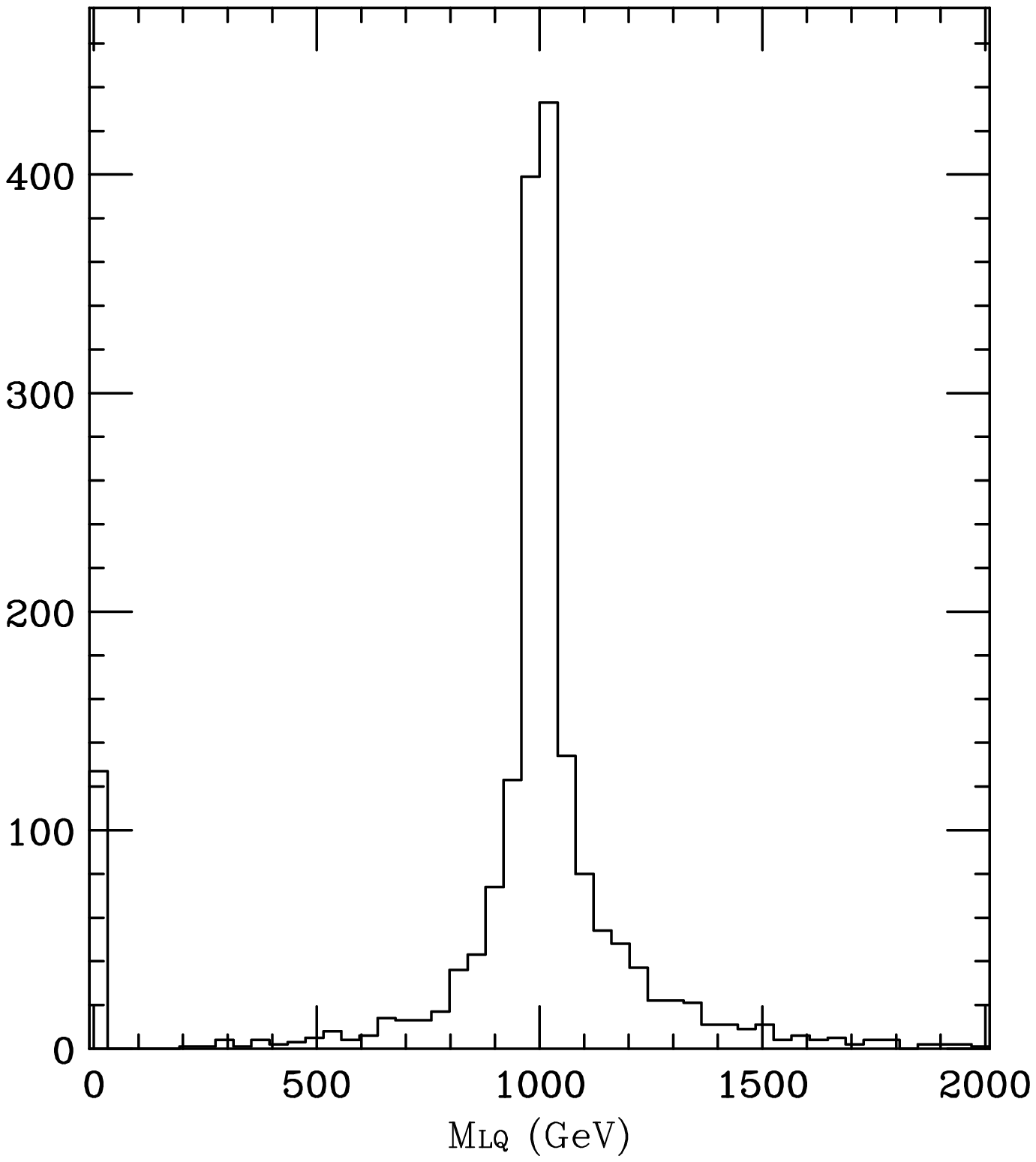}
\caption{Histograms of the solutions obtained at parton level for
  $S_0\bar{S}_0$ and decay to a semi-leptonic top, a hadronic top and
two tau leptons, for $M_{S_0} =   (0.25, 0.4, 1.0)$ TeV (from left to
right respectively). The zeroth bin contains the events where no real
solution has been found.}
\label{fig:mS0parton}
\end{figure}

We first perform the $(t\tau)(t\tau)$ analysis of the hard process (no initial or final state radiation, no underlying event) at parton-level without considering experimental or combinatoric effects, to examine its feasibility. For the
 majority of cases there are only two physical, approximately degenerate,
 solutions, which are close to the true leptoquark mass. The numerical
 solution of the quartic equation sometimes fails to yield roots. The
 results for true leptoquark masses $M_{S_0} = (0.25, 0.4, 1.0)$ TeV
 are shown in Figure~\ref{fig:mS0parton}, which includes histograms of the solutions
 obtained for $10^3$ events. The histogram includes a bin at
 0 where the events without solution are placed. These amount to about 10\% of the total events. At this level the reconstruction technique provides a good estimate of the leptoquark mass for all the trial true masses, lying within a few GeV of the true mass. 
\subsubsection{Experimental reconstruction}
We consider an $S_0$ leptoquark with mass $M_{S_0} = 400\gev$, for
which the cross section for production and decay into the topology of
Figure~\ref{fig:s0topology}, $S_0\bar{S}_0 \rightarrow t\tau
\bar{t}\bar{\tau} \rightarrow b \bar{b} jj \ell\nu \tau
\bar{\tau}$, is $\sigma = 0.187$ pb. The most significant background
in this scenario is $t \bar{t}$ production, with two extra jets misidentified as $\tau$s and subsequent decay of
the tops into $b \bar{b} jj \ell\nu$. The cross section for
this process is $242.4$ pb, overwhelming to begin with. There is also
potentially an irreducible $Ht\bar{t} \rightarrow \tau \bar{\tau} t
\bar{t}$ background which, for a Higgs of mass $M_H = 115 \gev$, has a
cross section of approximately 65 fb.  Since one of the main rejection
mechanisms is the reconstruction through the
solution of the kinematic equations, we do not expect this background to
contribute significantly. 

\begin{figure}[!htb]
  \centering
  \vspace{0.5cm}
  \includegraphics[scale=0.35, angle=270]{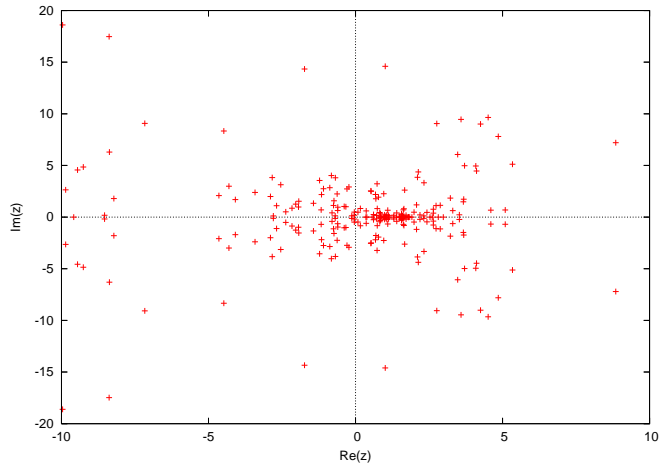}
  \vspace{0.75cm}
  \caption{The plot shows the complex values of the solutions for $z_2$ after solving the quartic equation for the events that have passed the experimental cuts. There exists a higher concentration of events about the positive real axis. The number of entries is 280 (4 solutions included for each of the 70 events).}
  \label{fig:zcomplex}
\end{figure}

\begin{figure}[!htb]
  \centering
  \vspace{0.75cm}
  \includegraphics[scale=0.40, angle=90]{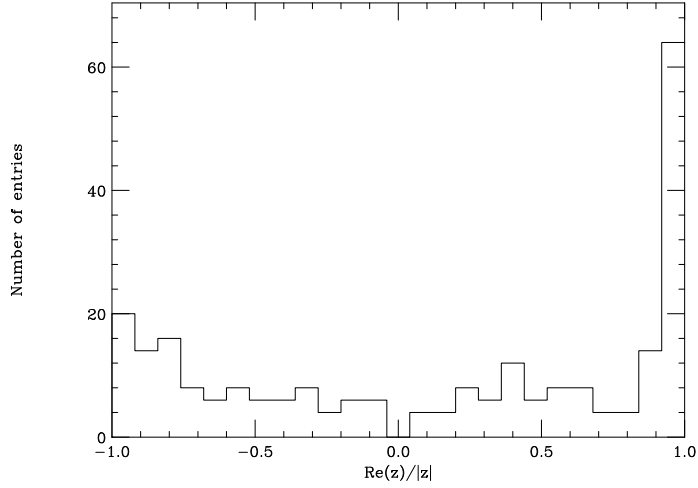}
  \vspace{0.75cm}
  \caption{The plot shows the ratio of the real part of $z_2$ and its modulus. The peak close to 1 demonstrates the clustering of the real positive solutions about the real axis and justifies the use of the real part as an input to the rest of the calculation. The number of entries is 280 (4 solutions included for each of the 70 events).}
  \label{fig:zcomplex2}
\end{figure}

\begin{figure}[!htb]
  \centering 
  \vspace{1.8cm}
  \hspace{6.0cm}
  \includegraphics[scale=0.55, angle=90]{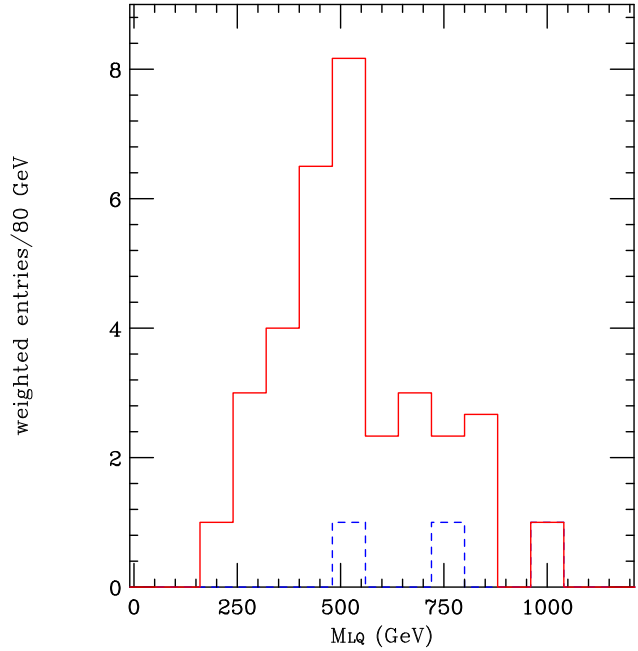}
  \vspace{0.5cm}
  \caption{Experimental reconstruction of the $S_0\bar{S}_0
    \rightarrow t\tau \bar{t}\bar{\tau} \rightarrow b \bar{b} jj
    \ell(=e,\mu) \nu \tau \bar{\tau}$ mode using the method described
    in the text.
  Note that each event has weight 1, distributed
    evenly amongst the solutions it yields. The signal is shown in
    red (35 entries) and the $t\bar{t}$ background in blue dashes (3 entries).}
  \label{fig:mS0ttauttau_btau22}
\end{figure}

We simulate the events with QCD initial state radiation
(ISR), final state radiation (FSR) and underlying event (UE). We use the default jet algorithm provided by the \texttt{Delphes}
package for the ATLAS configuration, anti-kt with $R=0.7$. We then
demand a set of relatively loose cuts on the full $t\bar{t}$ and
$S_0\bar{S_0}$ samples, since in a real experiment we wouldn't be able
to separate the different decay modes of the top quark or $S_0$ leptoquark. The cuts applied are the following:
\begin{itemize}
\item{the existence of a lepton in the event, being either a muon or
    electron, with $p_{T,\ell} > 30\gev$.}
\item{a minimum of 6 jets.}
\item{the missing transverse momentum in the event, $\slashed{E_T} > 20
    \gev$.}
\item{two $\tau$-tagged jets, with the extra requirement that they
    both have $p_{T,\tau} > 30\gev$.}
\item{no jets tagged as both $b$- and $\tau$-jets simultaneously.}
\end{itemize}
We also require that the highest-$p_T$ lepton is at a distance $\delta
R > 0.1$ from the $\tau$-tagged jets, since electrons may create a
candidate in the jet collection as well as the lepton collection. The
analysis then breaks up into different branches according to the number of $b$-tagged jets in an event. 
\begin{itemize}
\item{\textit{two $b$-tagged jets:} we look for one or two further jets (with $p_T > 30\gev$) that form an invariant mass close to the top mass, within $20 \gev$. One $b$-jet is then associated with the semi-leptonic top decay and the other with the hadronic top decay.}
\item{\textit{one or no $b$-tagged jets:} when there is one $b$-tagged jet we check whether it will satisfy the top mass conditions with any other (one or two) remaining jets, otherwise we associate it with the semi-leptonic top. If so, we look for any two or three jets that satisfy the top mass conditions, and form the hadronic top within a $20\gev$ mass window. For the remaining $b$-jets (or if there are no $b$-jets) we look for the remaining highest-$p_T$ jets. Any jets that are found in this way and called $b$-jets are required to have a $p_{T,b} > 30\gev$.} 
\end{itemize}
No solutions are found in the sample of 70 signal events passing the cuts, if we require the ratios $z_i$ to be purely real. Hence, the solutions to the quartic equation for the momentum ratios $z_i$,
described in Section~\ref{sec:ttauttaurecon}, are now allowed to be
complex in order to provide some signal, since even true
leptoquark events are smeared and distorted by detector and
QCD effects. We use the real part of $z_2$ as an input to the
calculation of the rest of the kinematic variables. This is reasonable since the experimental effects are expected to `smear' the position of the true value of $z_2$ in all directions in the complex plane. The effect is shown in Fig.~\ref{fig:zcomplex}, where we plot the real and imaginary parts of $z_2$ for the events that have passed the kinematic cuts. Evidently, there is a concentration of true solutions around the positive real axis, an effect exemplified by Fig.~\ref{fig:zcomplex2}, where we show the ratio of the real part of $z_2$ and its modulus. We have further demanded that the resulting momentum fractions are physical: $\mathcal{R}(z_{1,2}) > 1$, resulting in only real solutions for $M_{S_0}$.
Figure~\ref{fig:mS0ttauttau_btau22} shows a reconstruction plot for
leptoquarks of mass $M_{S_0} = 400\gev$. Note that each event was given weight 1, distributed evenly amongst the solutions it yields. In the case of complex $z_2$, we assume there is one solution corresponding to the complex conjugate pair.

Although the cuts applied are relatively weak, most of the rejection
comes from the requirement of two $\tau$-tagged jets. The background does not produce solutions in the
physical region often enough to be significant.

\subsection[$(q\nu)(q\nu)$ decay modes]{\boldmath $(q\nu)(q\nu)$ decay modes}
We can obtain the mass of the leptoquarks when both of them decay into
$b\nu$ or $t\nu$ using the $M_{T2}$ variable (Eq.~(\ref{eq:mt2})). Examples of these
decay mode are  $S_0\bar{S}_0 \rightarrow \bar{b}\bar{\nu} b\nu$ and
$\bar{S}^{(-)}_{1/2}S^{(-)}_{1/2} \rightarrow \bar{t}\nu t\bar{\nu}$.  

\subsubsection{Parton-level reconstruction}
\begin{figure}[!htb]
  \centering 
  \vspace{1.5cm}  
  \hspace{4.0cm}
    \includegraphics[scale=0.60, angle=90]{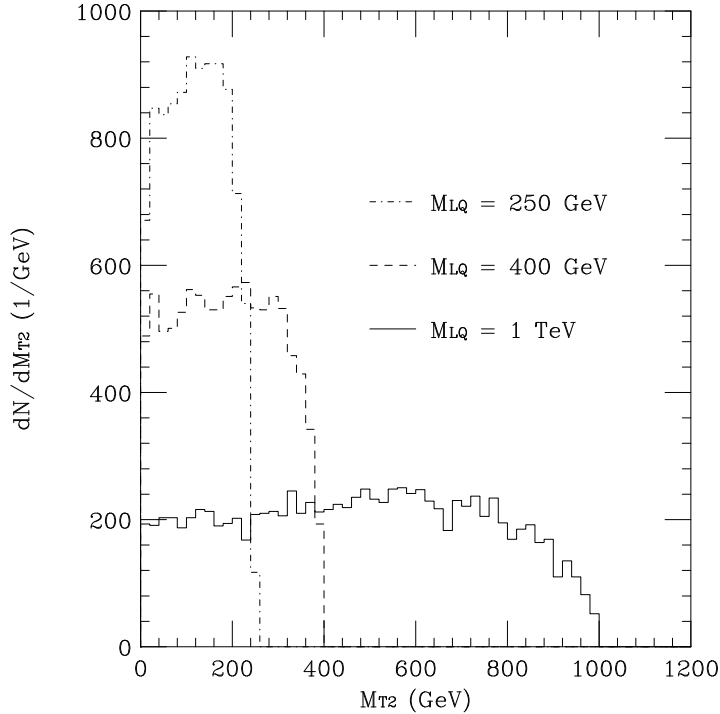}
  \vspace{0.5cm}  
\caption{The parton-level $M_{T2}$ distribution constructed for the $S_0\bar{S}_0 \rightarrow \bar{b}\bar{\nu} b\nu$ using the true $b$-quark momenta, for $M_{LQ} = 1, 0.4, 0.25$ TeV.}
\label{fig:bnubnumt2_parton}
\end{figure}

At parton level, the $(t\nu)(t\nu)$ and $(b\nu)(b\nu)$ decay modes are similar and hence we consider only the latter here. We first construct the $M_{T2}$ variable using the parton-level $b$-quark 4-momenta, in the absence of any experimental effects, ISR or FSR. The result is shown in Figure~\ref{fig:bnubnumt2_parton} for $M_{LQ} = (0.25, 0.4, 1)$ TeV, confirming the expected sharp edge in these idealised conditions.
\subsubsection{Experimental reconstruction}
\begin{figure}[!htb]
  \centering 
  \vspace{1.8cm}
  \hspace{3.0cm}
  \includegraphics[scale=0.40, angle=90]{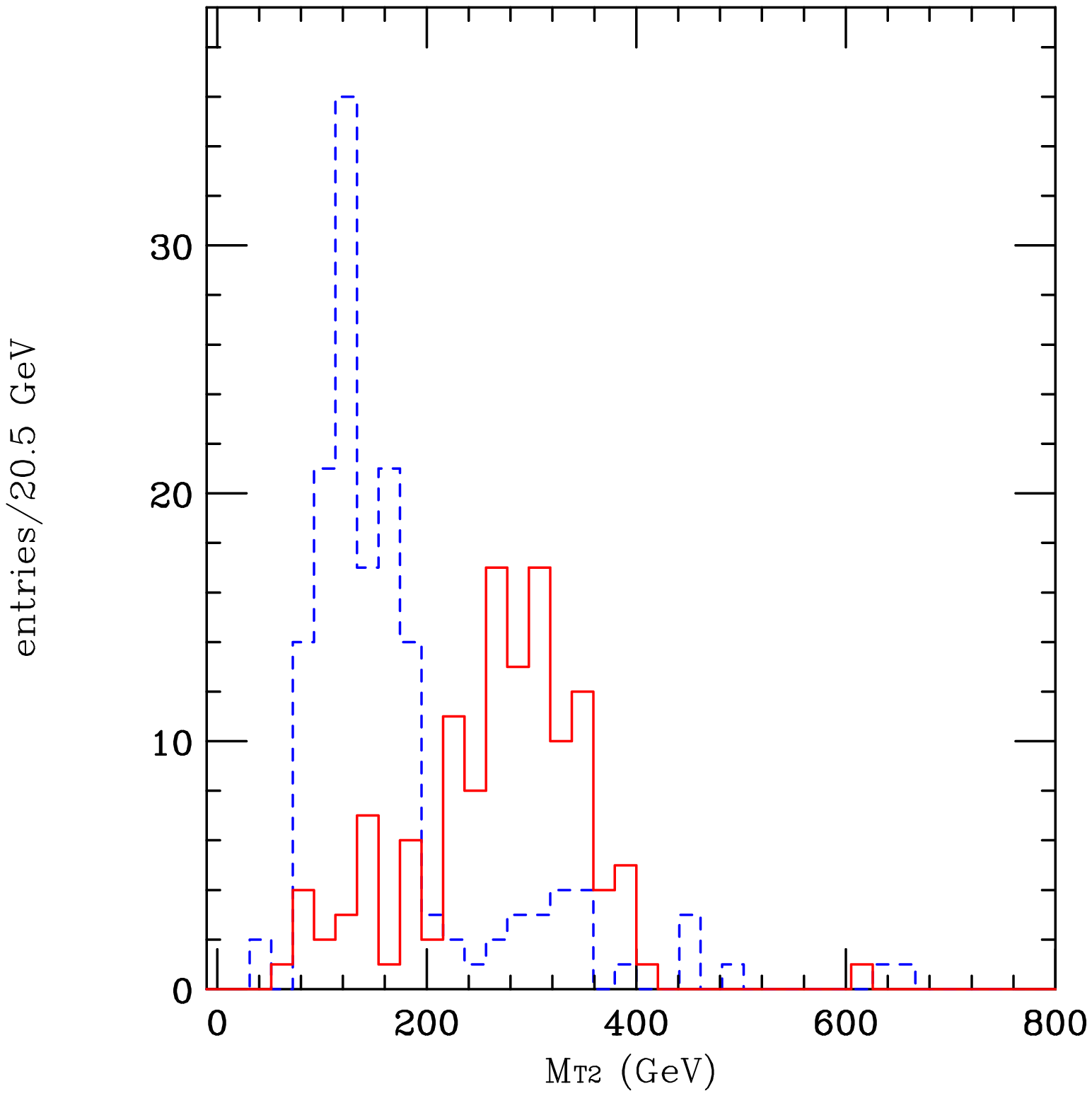}
  \hspace{5.0cm}
  \includegraphics[scale=0.40, angle=90]{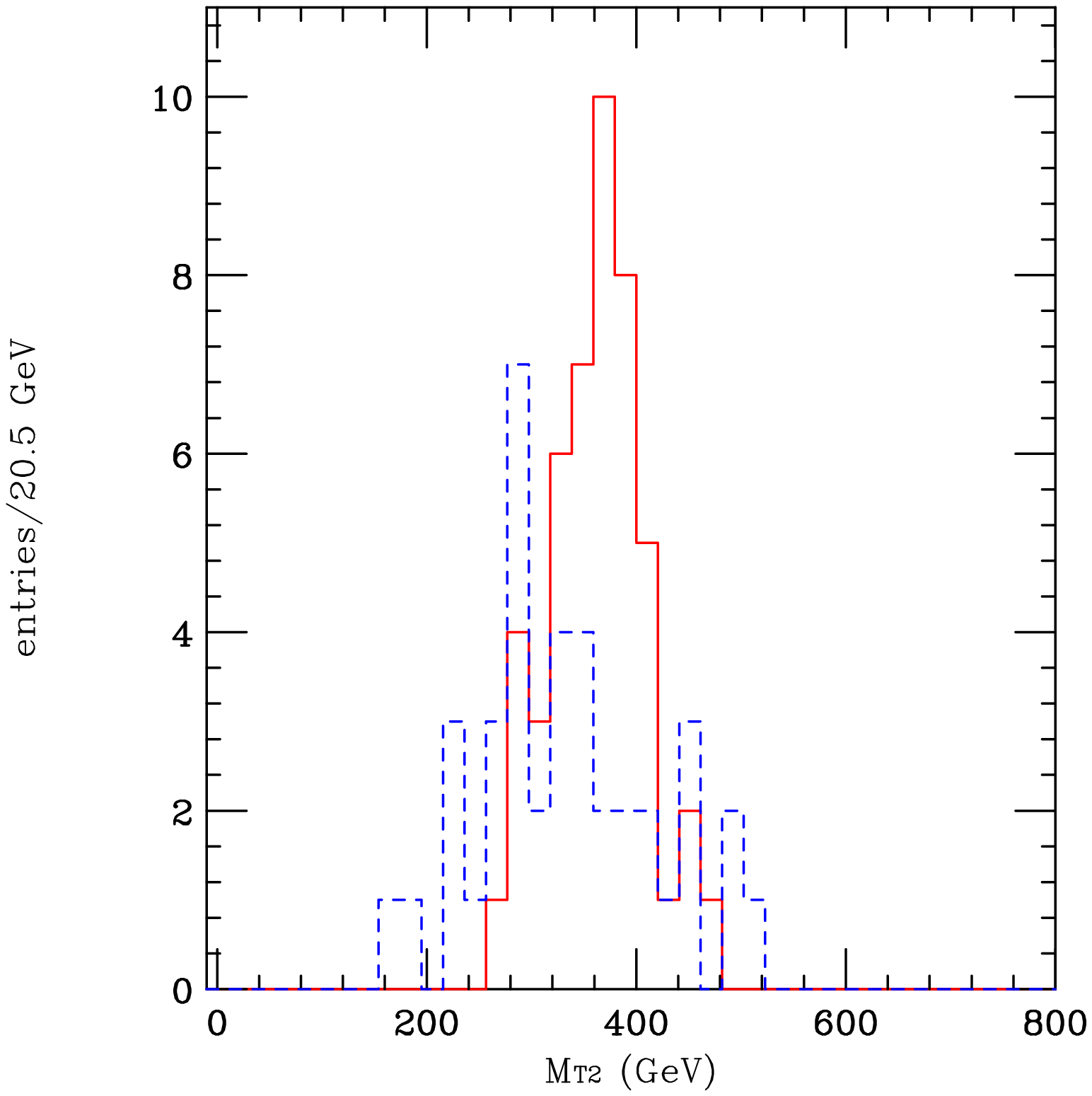}
  \vspace{0.5cm}
  \caption{Experimental reconstruction of the $S_0\bar{S}_0
    \rightarrow b\nu_\tau \bar{b} \bar{\nu_\tau}$ mode (left, 121 background
    events, 125 signal events)
    and $\bar{S}^-_{1/2}S^-_{1/2} \rightarrow \bar{t}\nu t\bar{\nu} $
    mode (right, 39
    background events, 48 signal events) using $M_{T2}$. ISR and FSR and the underlying event have been
    included in the simulation. The signal is given in
    red and the $t\bar{t}$ background in blue dashes.}
  \label{fig:mt2bnu}
\end{figure}

As before, we use the \texttt{Delphes} framework to simulate the
detector effects, with the settings stated in
Section~\ref{sec:outlinestrategy}. We demand two $b$-tagged jets in
both the $q = b$ and $q = t$ cases. In the latter we search for
combinations of 1 or 2 jets with the $b$-tagged jets which form the
top mass within a window of $10 \gev$.  We now require the following
cuts for the $(b\nu)(b\nu)$ case, on the full $S_0 \bar{S}_0$ sample:
\begin{itemize}
\item{two b-tagged jets with $p_{T,b} > 120 \gev$ each.}
\item{no electrons or muons in the event.}
\item{Missing transverse energy $\slashed{E_T} > 250 \gev$.}
\end{itemize}
For the $(t\nu)(t\nu)$ case we require the following cuts on the $\bar{S}^{(-)}_{1/2}S^{(-)}_{1/2}$ sample:
\begin{itemize}
\item{two b-tagged jets with $p_{T,b} > 80 \gev$ each.}
\item{no electrons or muons in the event.}
\item{Missing transverse energy $\slashed{E_T} > 260 \gev$.}
\end{itemize}
The resulting $M_{T2}$ distributions for the signal (blue) and $t\bar{t}$
background (red) can be seen in Figure~\ref{fig:mt2bnu}.
The $(t\nu)(t\nu)$ mode appears to be more challenging to reconstruct than the
$(b\nu)(b\nu)$ mode. This is due to the fact that the $t\bar{t}$ background is very similar to the
signal and the difficulties that are presented in reconstructing hadronic tops. Nevertheless, as the results show, it may
be possible to observe an excess over the $M_{T2}$ distribution of the
background and provide an estimate of the mass.
\subsection[$(q'\tau)(q \nu)$ decay modes]{\boldmath $(q'\tau)(q \nu)$ decay modes}
\begin{figure}[!t]
  \centering 
    \includegraphics[scale=0.60]{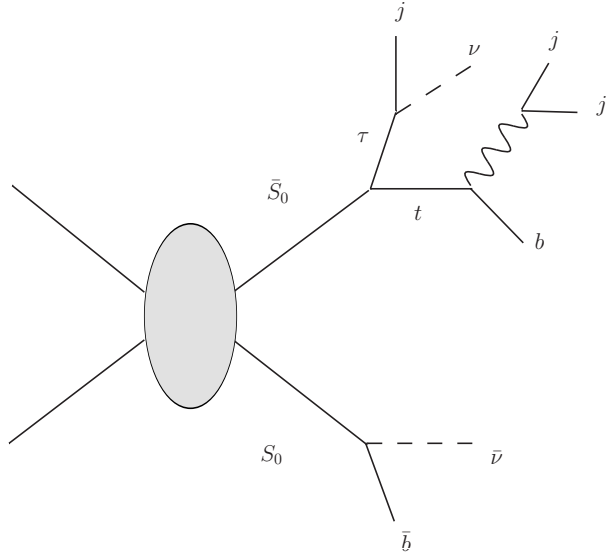}
 \caption{Pair production of $S_0$ leptoquarks with decay to $(t\tau)(b\nu)$, followed by hadronic top decay.}
\label{fig:topo-bntt}
\end{figure}
One possible event topology for the $S_0 \overline S_0 \to \bar b \bar \nu t \tau^-$ processes
is shown in Figure~\ref{fig:topo-bntt}.
When the top decays hadronically, the system has two neutrinos, 
one from an $S_0$ decay and another from a $\tau$ decay. 
If we can reconstruct the hadronic top correctly,
we can simply use $M_{T2}$ to obtain the mass of the leptoquarks. Similar topology is present in the $\bar{S}^{(-)}_{1/2} S^{(-)}_{1/2} \rightarrow (b\bar{\tau}) (\bar{t}\nu)$ decay mode.

It is known that the information from $M_{T2}$ is the same as that from
the `minimal kinetic constraints', in events where 
two identical particles decay to missing particles with the same mass~\cite{Serna:2008zk,Cheng:2008hk,Barr:2009jv}.
As discussed in Section\,\ref{sec:tttt}, in this type of event,
we can take advantage of the fact that, to a good approximation,
 the neutrino from a $\tau$ decay is travelling almost collinearly to the $\tau$ jet in the lab frame.
By including this constraint, we can define kinematical variables, 
$M_{\rm min}$ and $M_{\rm min}^{\rm bal}$, which perform better than $M_{T2}$ at parton-level,
as we will show in the following subsections.

\subsubsection{Kinematic reconstruction}\label{sec:ttaubnurecon}

In the above approximation we can write (neglecting masses):
\beq\label{eq:wpj}
p_{\nu_\tau} = w p_j~~~(w>0)\,.
\eeq
The second neutrino comes directly from the $S_0$ decay associated with a $b$-jet.
The transverse components of the momentum of this neutrino are constrained by
\beq
{\bf p}_{\nu} = {\bf p}_{\rm miss} - w {\bf p}_j\,.
\eeq  
There are two unknown parameters left, $w$ and $p_{\nu}^z$.  In terms of these,
we define two invariant mass variables:
\beq
m_{t \tau}^2(w) = (p_t + (1+w)p_j)^2 = m_t^2 + 2 (1+w) p_t\cdot p_j
\eeq
and
\beqn\label{eq:mbnu}
&& m_{b \nu}^2(w,p_{\nu}^z) \,=\, (p_b + p_{\nu})^2 \nonumber \\
&& ~~~=\, 2 E_b \sqrt{(\pmiss - w {\bf p}_j )^2 + (p_{\nu}^z)^2} - 2 {\bf p}_b \cdot (\pmiss -w {\bf p}_j)
- 2 p_b^z p_{\nu}^z\;\;.
\eeqn
Note that $m_{t \tau}$ does not depend on $p_{\nu}^z$ and is
 a monotonically increasing function of $w$ because $p_t\cdot p_j > 0$. 
We can now define two $M_{T2}$-like variables: 
\beq
M_{\rm min} = \min [ \max \{ m_{t\tau}, m_{b\nu}  \} ] \geq M_{T2}\;\;,
\eeq
and
\beq
M_{\rm min}^{\rm bal} = \min_{m_{t\tau} = m_{b\nu} } [m_{b\nu}]\;\;,
\eeq
where minimisation is taken for all possible ($w$, $p_{\nu}^z$).
By construction, both these quantities have an upper bound equal to the leptoquark mass:
\beq
M_{S_0} \ge M_{\rm min}, ~~~~M_{S_0} \ge M_{\rm min}^{\rm bal}\;\;.  
\eeq
Furthermore we show in Appendix~\ref{app:mminbal} that
\beq
M_{\rm min}^{\rm bal} \ge M_{\rm min}\;\;.
\label{eq:mb.gt.m}
\eeq
\subsubsection{Parton-level reconstruction}\label{sec:partonqtauqnu}
\begin{figure}[!t]
 \begin{center}
  \hspace{3.0cm}
  \includegraphics[width=4.5cm, angle = 90]{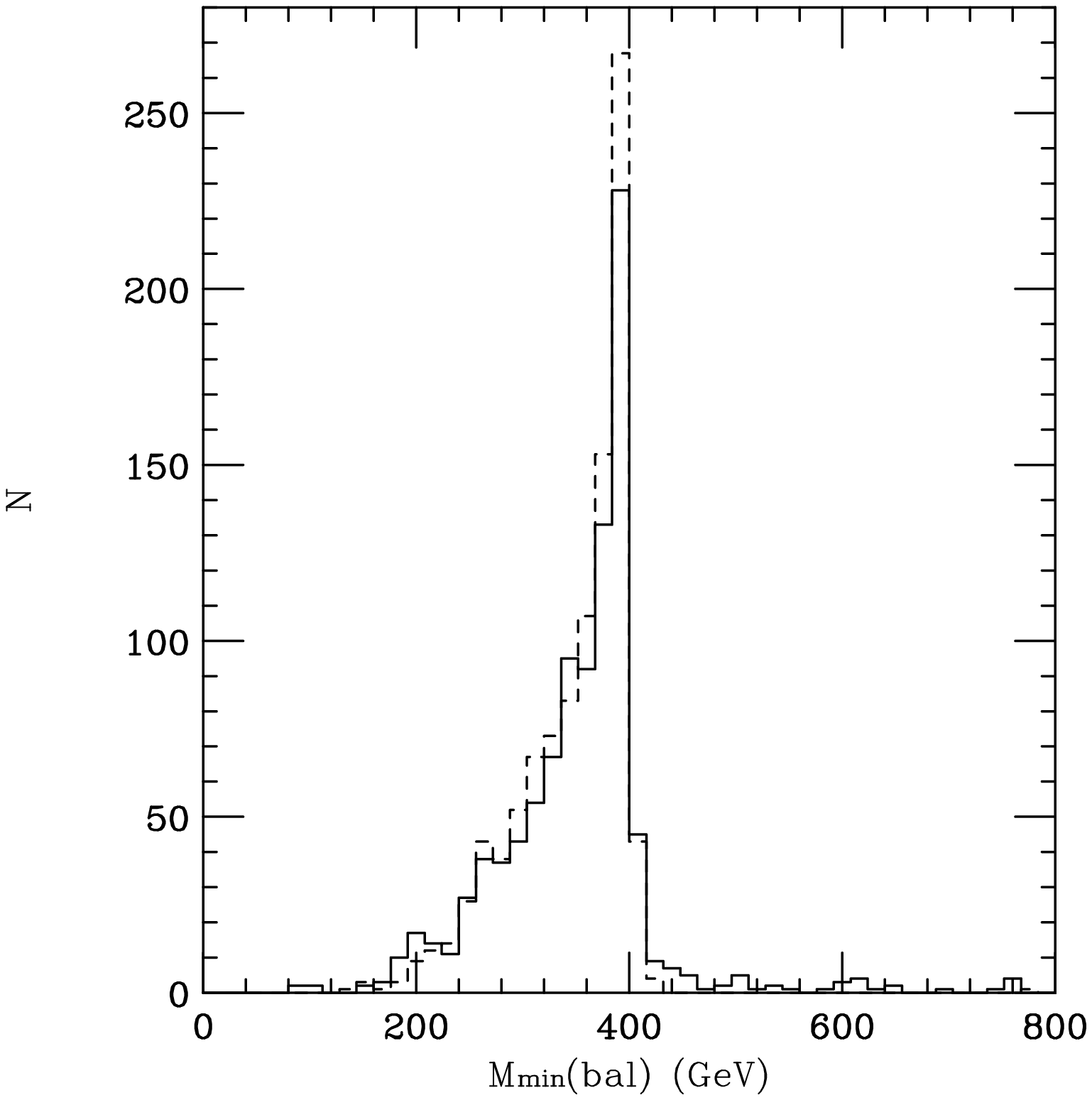}
  \hspace{3.5cm}
  \includegraphics[width=4.5cm, angle = 90]{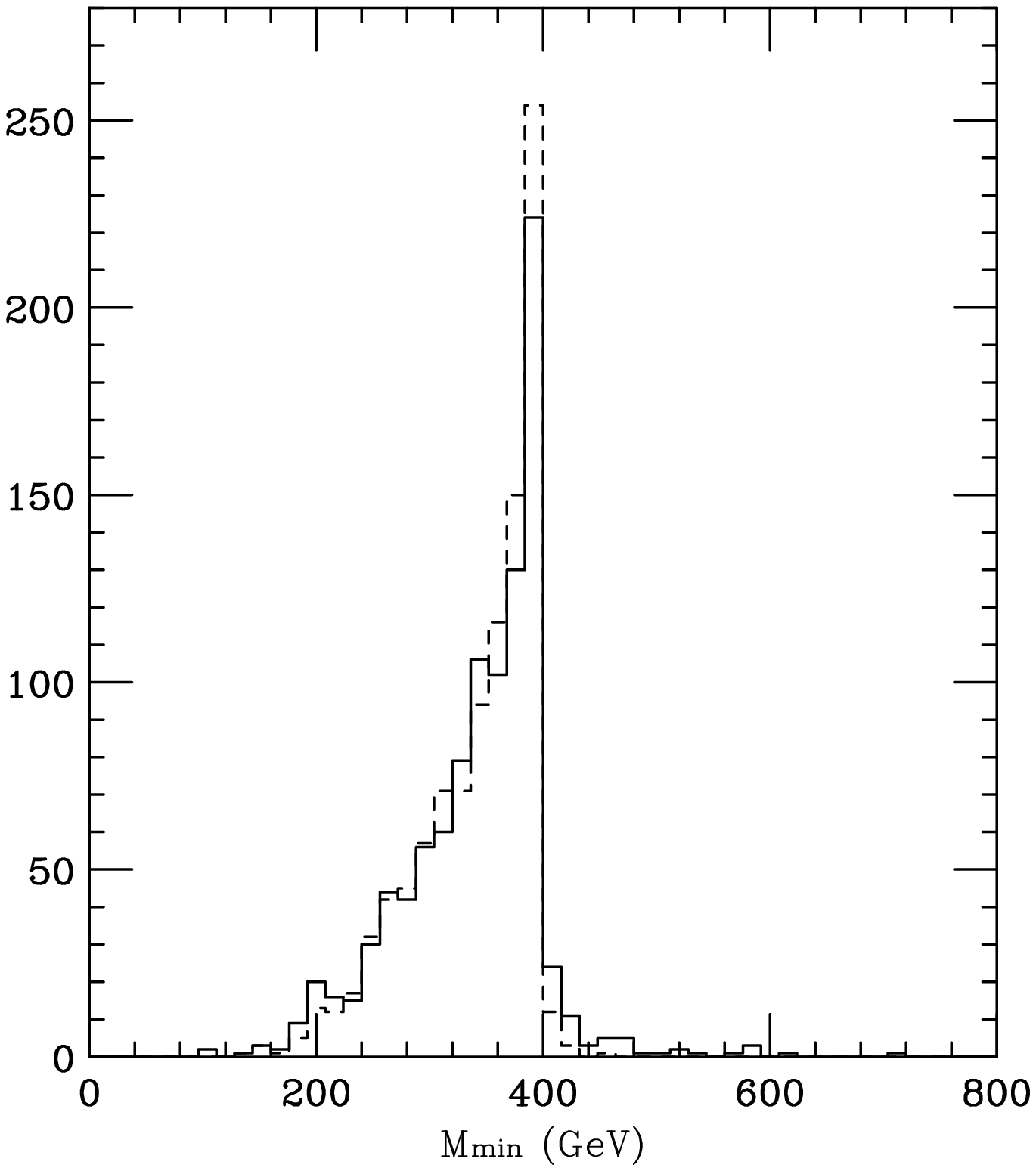}
  \hspace{3.5cm}
  \includegraphics[width=4.5cm, angle = 90]{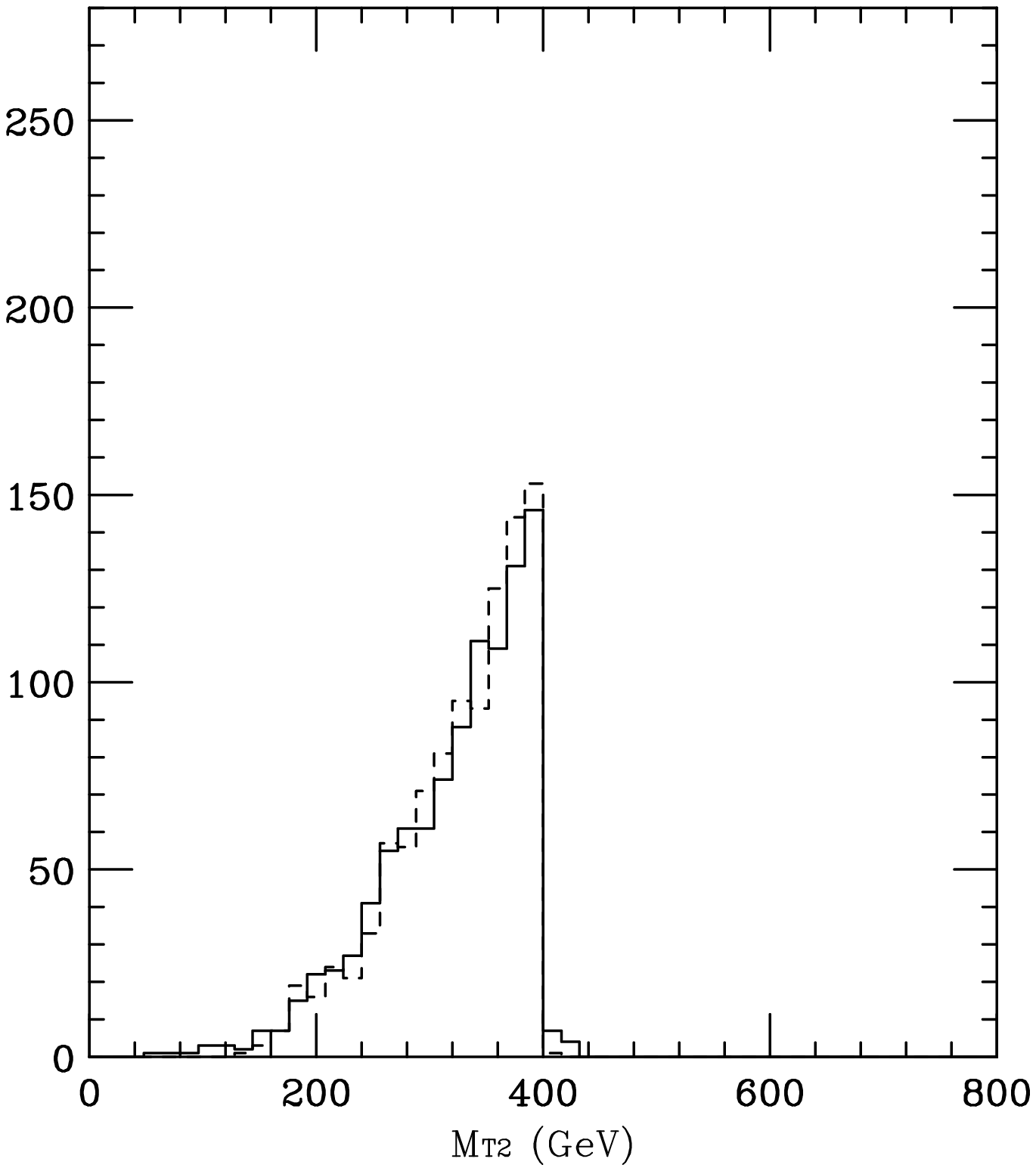}
 \caption{ Parton level distributions of $M_{\rm min}^{\rm bal}$ (left), $M_{\rm min}$ (centre) 
 and $M_{T2}$ (right) for $(b\nu)(t\tau)$ (solid curve) and $(t\nu)(b\tau)$ (dashed curve).
 }
\label{fig:mmin}
\end{center}
\end{figure}
Figure\,\ref{fig:mmin} shows the parton-level distributions of $M_{\rm min}^{\rm bal}$, $M_{\rm min}$ and $M_{T2}$.
Here we generated 1000 events and took only the true combination of the jet assignment.
As can be seen, all the distributions have clear edge structures at the input leptoquark mass of 400\,GeV.

\begin{figure}[!t]
 \begin{center}
      \vspace{1.0cm}
   \hspace{4.0cm}
   \includegraphics[width=4.5cm, angle = 90]{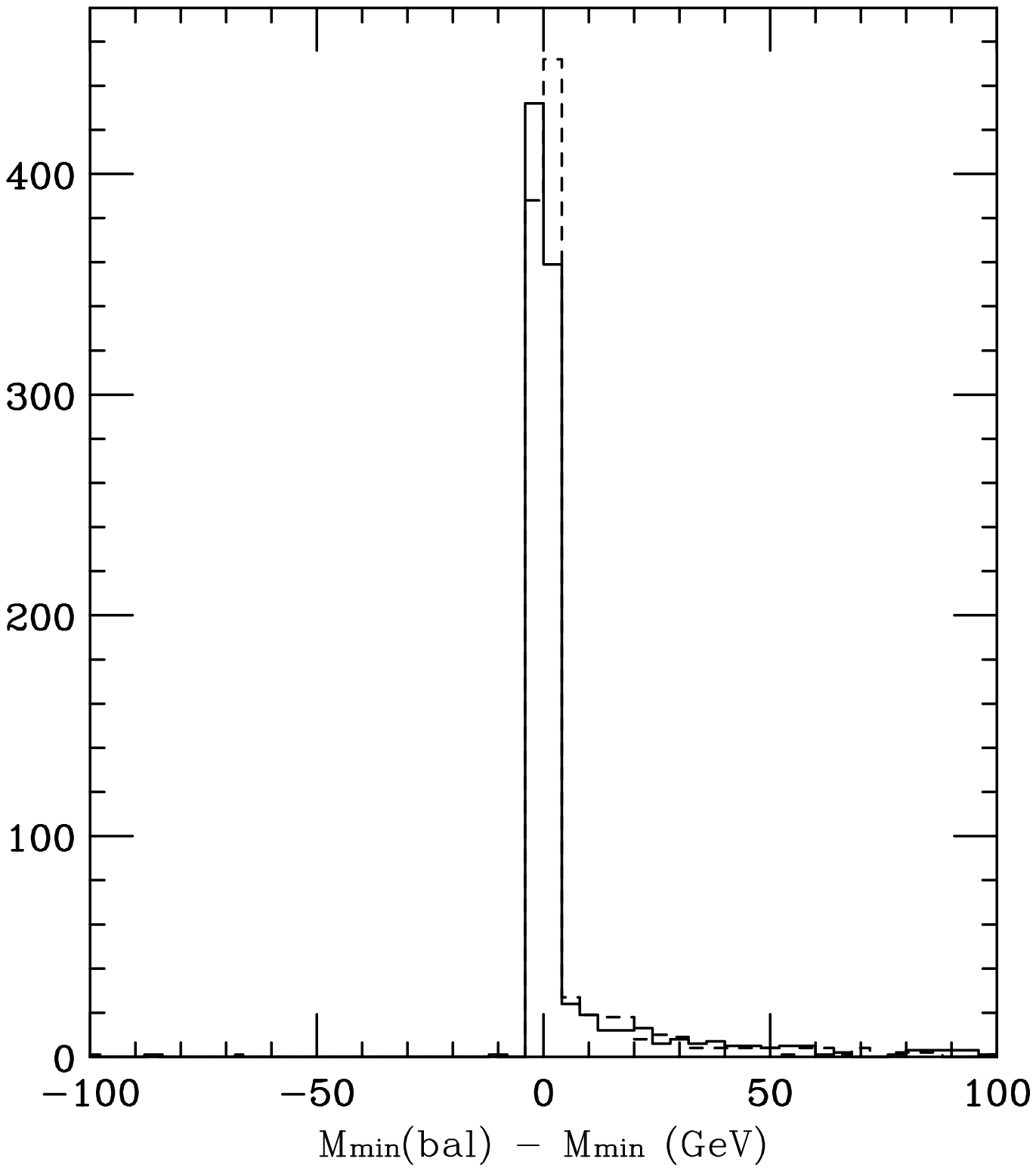}
  	\hspace{4.0cm}
  \includegraphics[width=4.5cm, angle = 90]{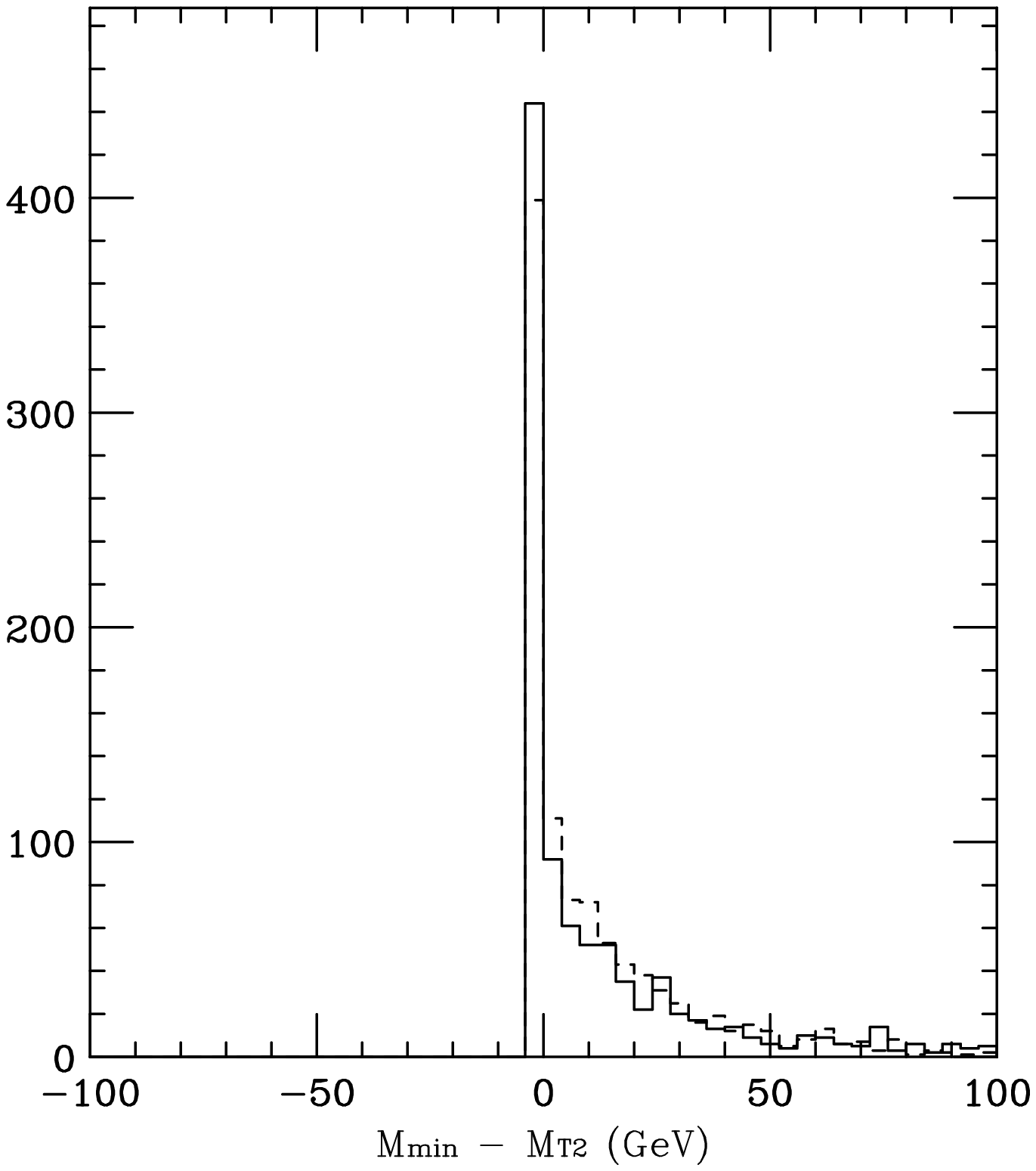}
 \caption{
 Parton level
 distributions of $M_{\rm min}^{\rm bal}-M_{\rm min}$ (left) and 
 $M_{\rm min}-M_{T2}$ (right) for $(b\nu)(t\tau)$ (solid curve) and $(t\nu)(b\tau)$ (dashed curve)}
\label{fig:diff}
\end{center}
\end{figure}

In order to compare these variables we took the differences, shown in Figure\,\ref{fig:diff}. 
The relation $M_{\rm min}^{\rm bal} \ge M_{\rm min} \ge M_{T2}$ is seen to
hold on an event-by-event basis.  
This implies that $M_{\rm min}^{\rm bal}$ and $M_{\rm min}$ are more powerful than $M_{T2}$
for determining the mass of the leptoquark, at least at parton level.

\subsubsection{Experimental reconstruction}
The settings for experimental reconstruction used for the
\texttt{Delphes} fast simulation remain unaltered in the
present analysis (see
Section~\ref{sec:outlinestrategy}). We apply the following event
selection cuts to the full $S_0 \bar{S_0}$ signal and the $t\bar{t}$
background:
\begin{itemize}
\item{at least four jets found in each event.}
\item{exactly one $\tau$-tagged jet with $p_T > 120 \gev$.}
\item{no, one or two $b$-tagged jets with $p_T > 60 \gev$.}
\item{missing transverse energy, $\slashed{E_T} > 200 \gev$.}
\end{itemize}
For the $b$-jet originating from the leptoquark decay,
we choose the highest-$p_T$ $b$-tagged jet when there are
two $b$-tagged jets and the highest-$p_T$ jet (excluding the
$\tau$-tagged jet) when there are no $b$-tagged jets. We use all the
remaining jets with $p_T > 30\gev$, (not identified as the $b$-jet from the leptoquark) to
search for one, two or three jets that form an invariant mass close to the top
mass, within a $20 \gev$ window. We apply the
additional constraint that the difference between the $p_T$ of the $\tau$-tagged jet and the $p_T$ of the $b$-tagged jet, $p_{T,\tau} - p_{T,b} > -
10 \gev$. This eliminates a high fraction of the $t\bar{t}$ background
since the $\tau$s in that sample originate from the $W$ decay and are
expected to have lower $p_T$ on average than the $b$s that originate
directly from the top. On the contrary, in the leptoquark signal the
$\tau$ and $b$ transverse momenta are expected to be more equal on average. 

\begin{figure}[!t]
 \begin{center}
   \vspace{1.0cm}
  \hspace{4.0cm}
  \includegraphics[width=4.5cm, angle = 90]{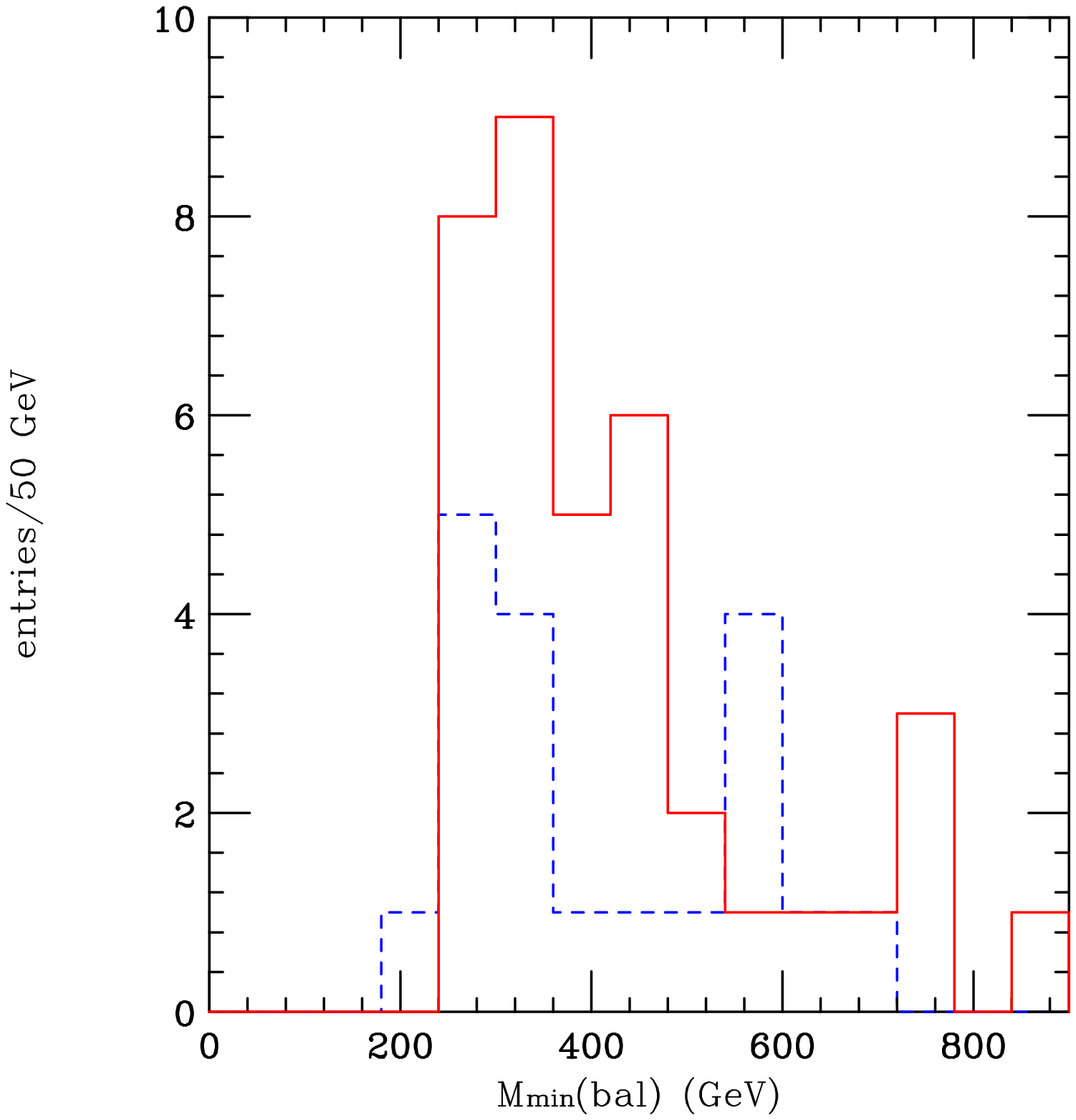}
  \hspace{4.0cm}
  \includegraphics[width=4.5cm, angle = 90]{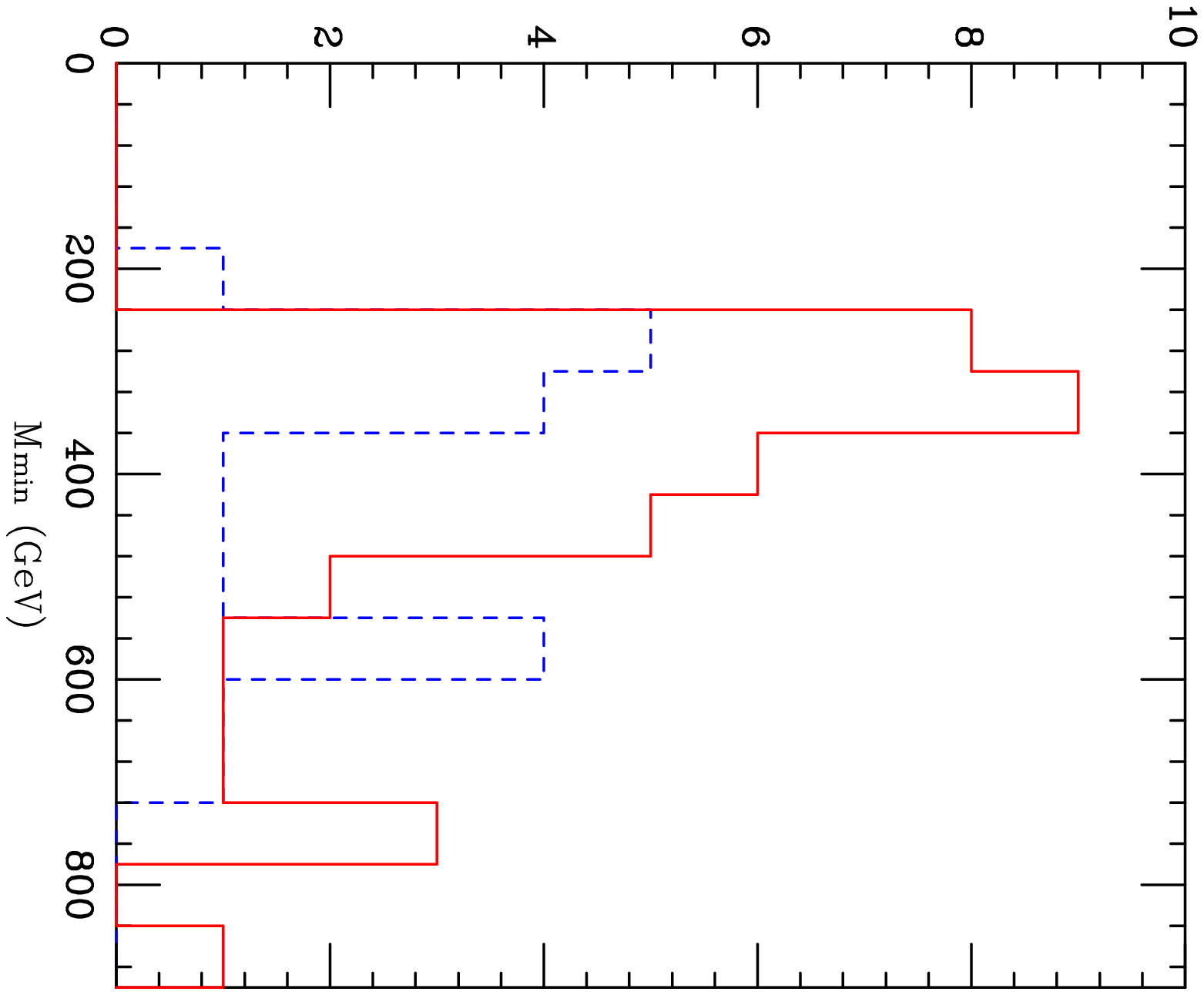}
  \hspace{4.0cm}
  \includegraphics[width=4.5cm, angle = 90]{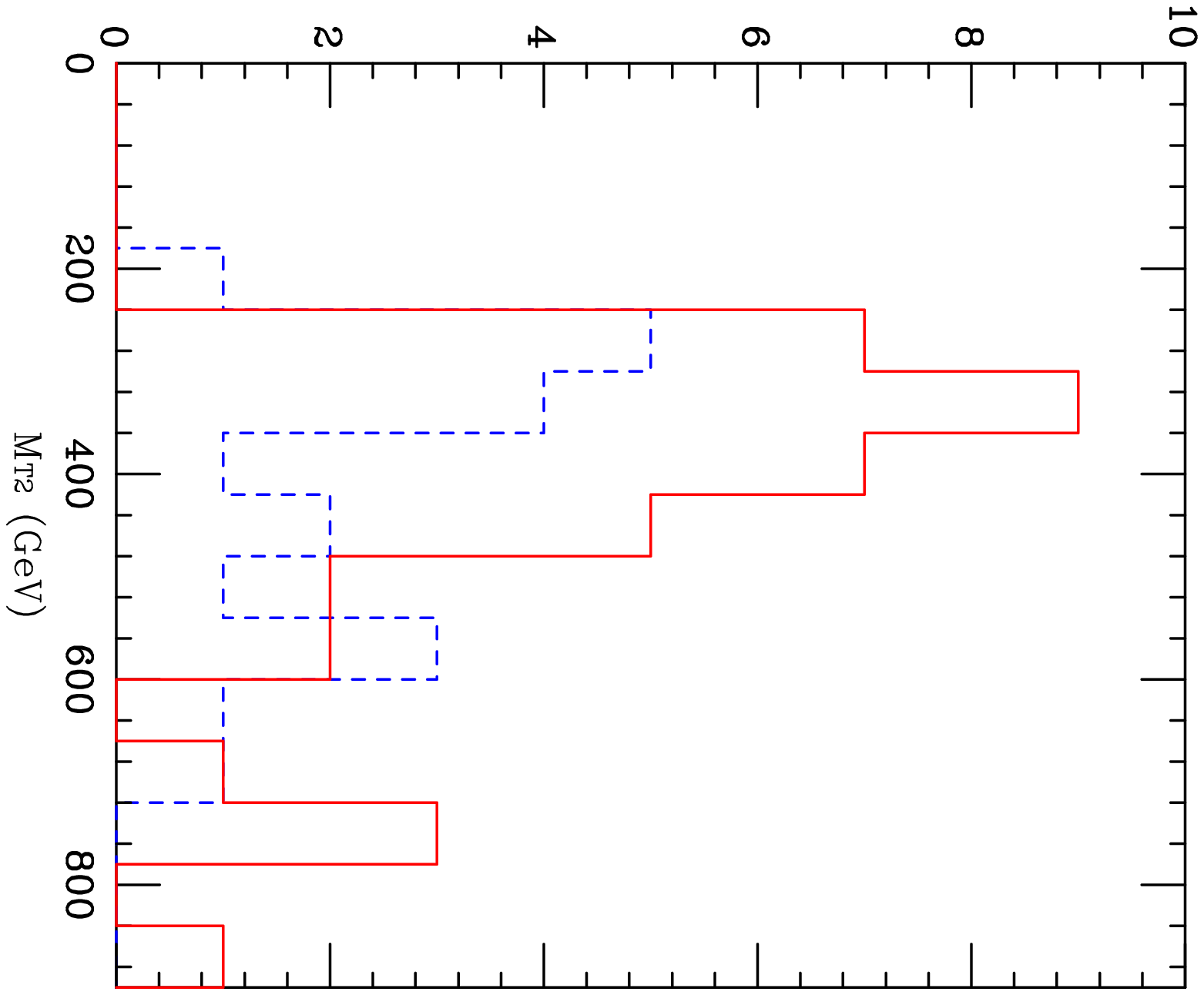}
  \caption{Reconstructed distributions $M_{\rm min}^{\rm bal}$ (left),
   $M_{\rm min}$ (centre) 
 and $M_{T2}$ (right) for $(b\nu)(t\tau)$ signal (red) and  $t\bar{t}$ background
 events (blue dashes) including ISR, FSR and the underlying event.  There are
37 signal events and 19 background events in all plots.}
\label{fig:bnuttaures}
\end{center}
\end{figure}

The resulting distributions are shown in Figure~\ref{fig:bnuttaures}. Due to the low
number of events passing the selection cuts, it is not obvious whether the
$M_{\mathrm{min}}^{\mathrm{bal}}$ observable performs better than $M_{\rm min}$
and $M_{T2}$. We checked, however, that the three distributions
satisfy the same inequalities presented in Figure~\ref{fig:diff} for
the parton-level reconstruction.

\begin{figure}[!t]
 \begin{center}
   \vspace{1.0cm}
  \hspace{4.0cm}
  \includegraphics[width=4.5cm, angle = 90]{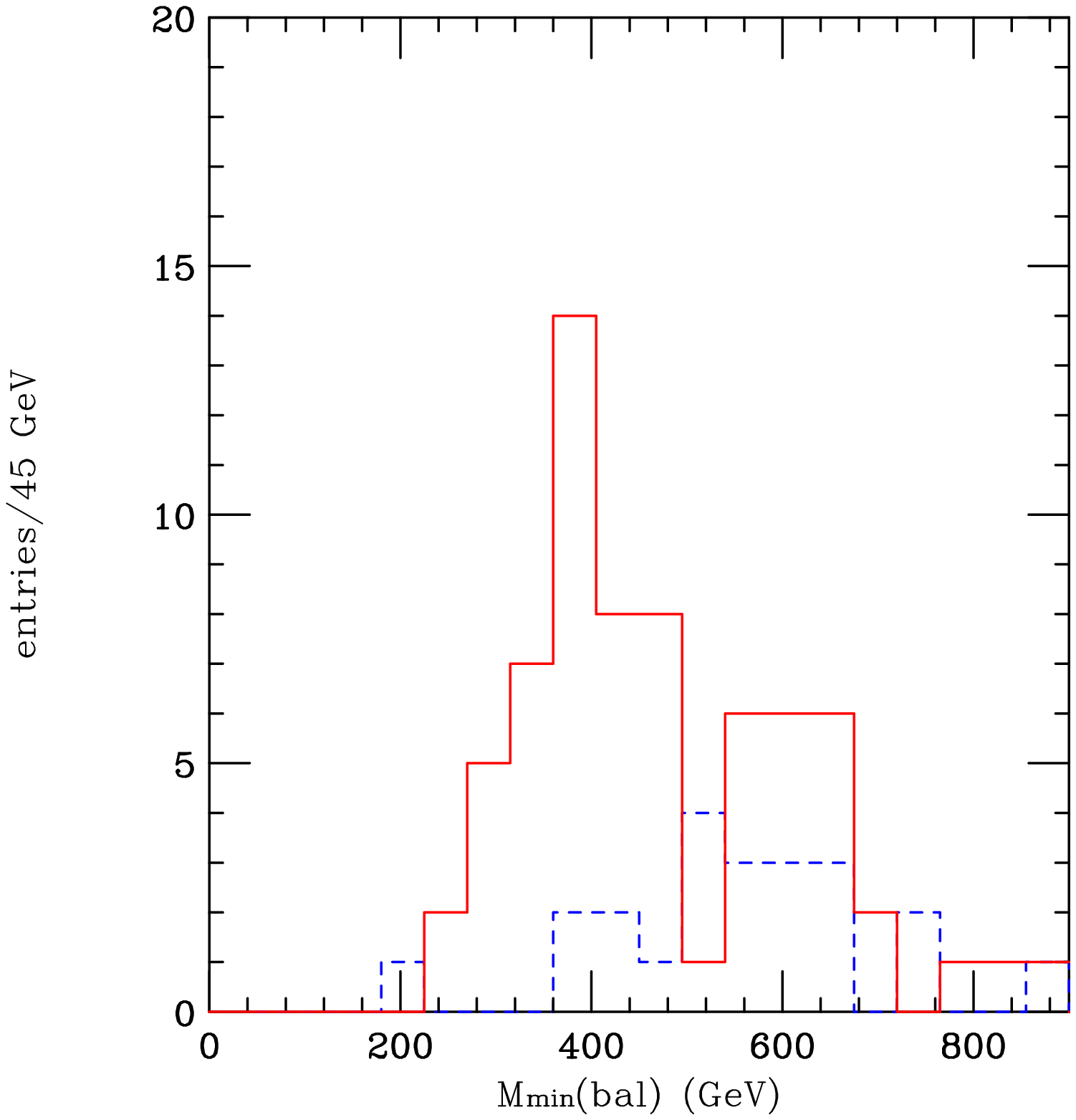}
  \hspace{4.0cm}
  \includegraphics[width=4.5cm, angle = 90]{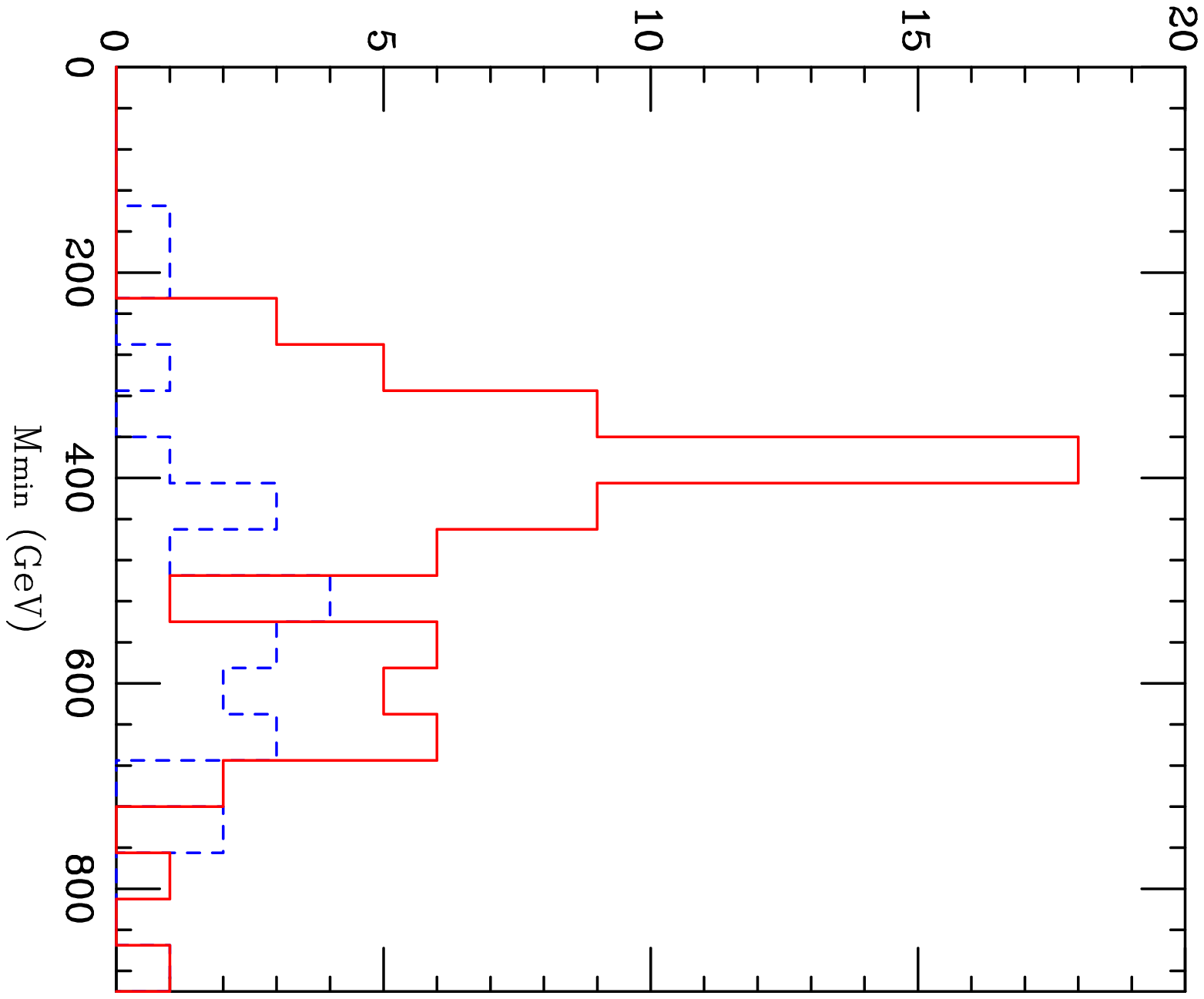}
  \hspace{4.0cm}
  \includegraphics[width=4.5cm, angle = 90]{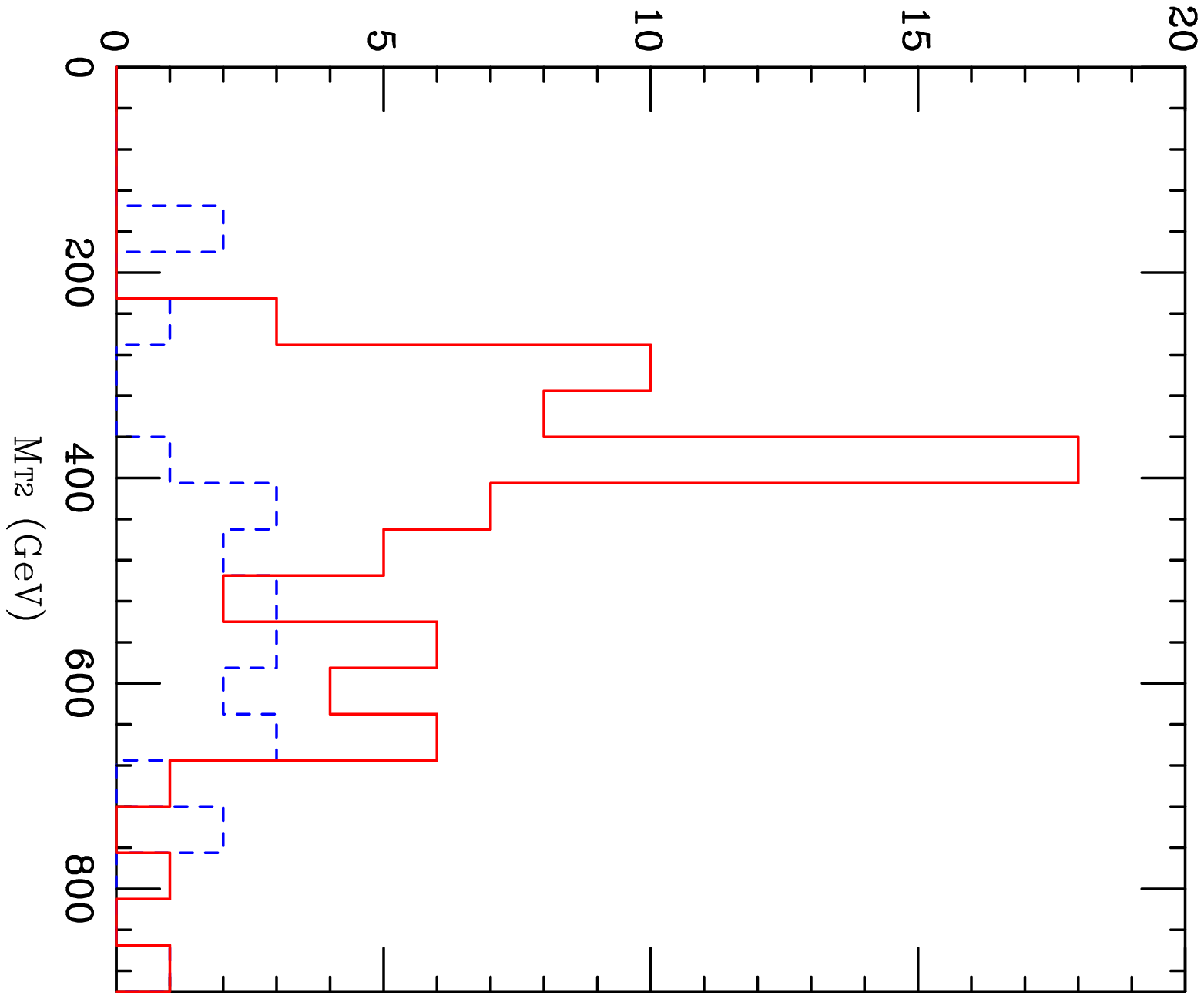}
  \caption{Reconstructed distributions $M_{\rm min}^{\rm bal}$ (left),
   $M_{\rm min}$ (centre) 
 and $M_{T2}$ (right) for the $(t\nu)(b\tau)$ signal (red) and $t\bar{t}$ background 
 events (blue dashes) including ISR, FSR and the underlying event.  There are
 68, 72 and 72 signal events (left to right) and 22, 23, 23 background events (left to right).}
\label{fig:tnubtaures}
\end{center}
\end{figure}

The reconstruction strategy for the $(t\nu)(b \tau)$ mode follows the technique described in Section~\ref{sec:ttaubnurecon} for the $(b\nu)(t \tau)$ case, with the simple replacement $b \leftrightarrow t$. The assignments of $b$-jets and top-jets is performed in the same way as in the $(b\nu)(t \tau)$ analysis, with the following cuts applied to the full $\bar{S}^{(-)}_{1/2} S^{(-)}_{1/2}$ sample:
\begin{itemize}
\item{at least four jets found in each event.}
\item{exactly one $\tau$-tagged jet with $p_T > 190 \gev$.}
\item{no, one or two $b$-tagged jets with $p_T > 40 \gev$.}
\item{missing transverse energy, $\slashed{E_T} > 120 \gev$.}
\end{itemize}
There is also a cut on the reconstructed hadronic top jet, of $p_T > 120\gev$ and that its invariant mass lies within $20\gev$ of the top mass. The results are shown in Figure~\ref{fig:tnubtaures}.  Note that the background that would be present due to the $S^{(+)}_{1/2}$ leptoquark has not been included.

Although at parton-level, the variable $M_{\mathrm{min}}^{\mathrm{bal}}$ performs better than $M_{\mathrm{min}}$ and $M_{T2}$, it seems to become unstable after including experimental errors, with some events failing to produce a value within the range of the plots shown in Figure~\ref{fig:tnubtaures}. The origin of the instability is the additional assumption of the leptoquark masses being equal, which is satisfied at parton-level (up to small width effects) but does not hold exactly after detector simulation. For the events for which no solution is found, we assign  $M_{\mathrm{min}}^{\mathrm{bal}} = M_{\mathrm{min}}$. Even after this readjustment, there are a few events for which a solution for $M_{\mathrm{min}}^{\mathrm{bal}}$ is found and lies outside the region shown. Therefore, $M_{T2}$ and $M_{\mathrm{min}}$ appear to be preferable as experimental observables.
\subsection[$(b\tau)(b\tau)$ decay mode]{\boldmath $(b\tau)(b\tau)$ decay mode}
\subsubsection{Kinematic reconstruction}\label{sec:btaubtaukin}
The $(b\tau)(b\tau)$ mode can be fully reconstructed if one again assumes collinearity of the $\tau$-jets and $\tau$-neutrinos: $p_{\tau,i} = z_i p_{j,i}$ ($i = 1,2$, $z_i>1$). This implies that the missing momentum from each $\tau$ can be written as $\slashed{p_i} = (z_i - 1) p_{j,i}$. Hence, we may write the following equalities for the components of the measured missing transverse momentum:
\begin{eqnarray}
p_{\rm miss}^x &=& p^x_{j1} ( z_1 - 1 ) + p^x_{j2} (z_2 - 1) \nonumber \\
p_{\rm miss}^y &=& p^y_{j1} ( z_1 - 1 ) + p^y_{j2} (z_2 - 1)\;\;.
\end{eqnarray}
The above equations may be written in matrix form and inverted to give:
\begin{eqnarray}\label{eq:z1z2btaubtau}
z_1 &=& 1+ \frac{ p^y_{j2} p^x_{\rm miss} - p^x_{j2} p^y_{\rm miss} } { p^x_{j1} p^y_{j2} - p^y_{j2} p^x_{j2} }\;\;, \nonumber \\
z_2 &=& 1 -\frac{ p^y_{j1} p^x_{\rm miss} - p^x_{j1} p^y_{\rm miss} } { p^x_{j1} p^y_{j2} - p^y_{j2} p^x_{j2} }\;\;.
\end{eqnarray}
Now the invariant mass of each of the two leptoquarks may be written as $m_S^2 = (p_b + p_\tau)^2$, resulting in the following expression:
\begin{equation}
m_S^2 = 2 z_i p_{bi} \cdot p_{ji}\;\;,
\end{equation}
where we have neglected the $\tau$ and $b$-quark mass terms. Using Eqs.~(\ref{eq:z1z2btaubtau}), we obtain two values of $m_S$ on an event-by-event basis. At parton level, with the correct jet assignments, these solutions approximate the leptoquark mass very closely, up to the collinearity approximation.

\subsubsection{Experimental reconstruction}
\begin{figure}[!htb]
  \centering 
  \vspace{1.8cm}
  \hspace{5.5cm}
  \includegraphics[scale=0.55, angle=90]{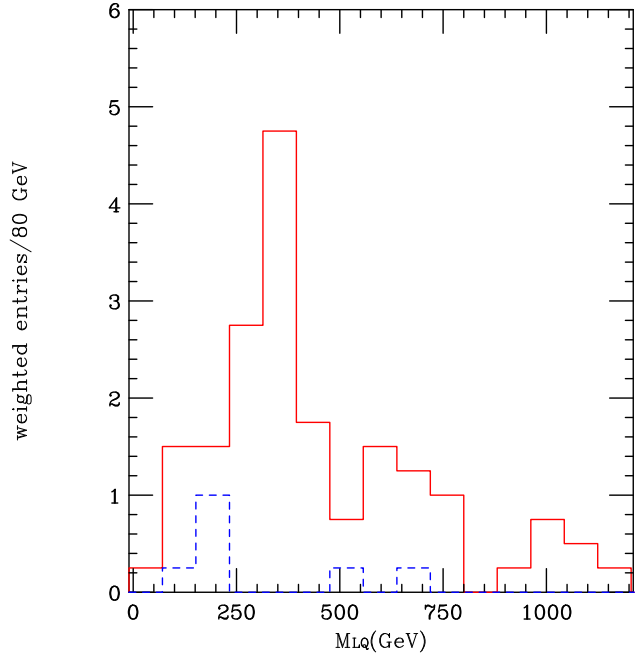}
  \vspace{0.5cm}
  \caption{Experimental reconstruction of the $(b\tau)(b\tau)$ mode using the method described
    in the main text. ISR and FSR have been
    included in the simulation. Note that each solution has weight 0.25. The signal is shown in
    red (18.75 entries) and the $t\bar{t}$ background in blue dashes (1.75 entries).}
  \label{fig:mbtaubtau}
\end{figure}

As before, the \texttt{Delphes} framework has been used with identical settings. The following cuts have been applied to the $S^{(+)}_{1} \bar{S}^{(+)}_{1} \rightarrow (\bar{b}\bar{\tau}) (b\tau)$ mode: 
\begin{itemize}
\item{at least 4 jets present in the event.}
\item{two $\tau$-tagged jets with $p_T > 140 \gev$.}
\item{missing transverse energy $\slashed{E_T} > 140 \gev$.}
\end{itemize}
We accept events with no, one or two $b$-tagged jets. If there are less than two $b$-jets,
we search for the highest-$p_T$ non-tagged jet(s) to obtain two $b$-jets. We apply a cut of $p_T > 50\gev$ on these. There are two possible assignments of the $b\tau$ combination, resulting in a total of four solutions. The resulting distribution for the mass solutions, as described in Section~\ref{sec:btaubtaukin}, is show in Figure~\ref{fig:mbtaubtau}.
The $t\bar{t}$ background appears to be under control, with 1.75 entries in the mass histogram. 

We also considered the $b\bar{b}jj$ background, for which we generated events using \texttt{Alpgen} v2.13~\cite{Mangano:2002ea}, applying the $p_T$ cuts on the four parton-level objects. We concluded that we can safely ignore this background since the expected number of events with two $\tau$-tagged jets was $\mathcal{O}(1)$, before applying any restrictions on the missing transverse energy, $\slashed{E_T}$. Note that the backgrounds to this decay channel originating from the other members of the leptoquark multiplet have not been included.
\subsection{Determination of quantum numbers}
In the ideal scenario where all of the decay modes of a leptoquark multiplet are seen, the quantum numbers can be deduced without ambiguity. For example if we only observe combinations of $(t\tau)$ and $(b\nu)$ decay modes, then the only likely candidate is an $S_0$ singlet. If in conjunction with these decay modes we observe $(b\tau)$ and $(t\nu)$ decay modes, with corresponding total rates, then we might guess that we have observed the $S_1$ multiplet. 

The issue is more complicated if some decay modes are missed.  For example if only the $(t\tau)(t\tau)$ decay mode has been seen, we might assume that we have observed the pair production of an $\tilde{S}'_0$ leptoquark. However, we might have observed the $(t\tau)(t\tau)$ decay of an $\tilde{S}_{1/2}'^{(+)}$ leptoquark pair and missed the more challenging $(t\nu)(t\nu)$ mode of the $\tilde{S}_{1/2}'^{(-)}$ leptoquarks. In this case we would need to examine the helicities and charges of the decay products: the $\tilde{S}_{1/2}'^{(+)}$ decays to $\bar{t}_L \bar{\tau}_L$ and $\bar{t}_R \bar{\tau}_R$ whereas the $\tilde{S}'_0$ decays to  $t_R \bar{\tau}_L$ and $t_L \bar{\tau}_R$.  Since we can reconstruct all decay products of the top and $\tau$ without combinatorial ambiguity, using measured leptoquark mass as an input, there is hope that we could measure top~\cite{Shelton:2008nq, Godbole:2006tq, Dalitz:1991wa} and $\tau$~\cite{Guchait:2008ar, Godbole:2008it} polarisations simultaneously. This would allow us to distinguish these two cases. We leave investigation of the feasibility of this to future work.

\section{Conclusions}\label{sec:conc}
If strongly-coupled dynamics solves the hierarchy problem of electroweak symmetry breaking, the question arises of how best to discover it at the LHC. Existing constraints coming from electroweak precision tests tell us that, at least at low energies, any model of strong dynamics must be a lot like the Standard Model (with perhaps a faint hope of observable deviations in the Higgs sector \cite{Low:2009di,Gripaios:2009pe}). 

Existing constraints coming from flavour physics are somewhat different in that, while the data are certainly consistent with the Standard Model, naturalness arguments suggest that strongly-coupled theories should differ from the Standard Model in the flavour sector. Indeed, fermion masses should arise via mixing between elementary and composite fermions of the strongly-coupled sector.

If that is so, then composite leptoquarks (or diquarks) may also appear, coupled predominantly to third-generation fermions. These would provide a spectacular signature at the LHC.  Their Standard Model quantum numbers imply that they would be produced strongly as conjugate particle-antiparticle pairs, decaying into third-generation quarks and leptons in the combinations summarised in Table~\ref{tb:strategy}.  We have proposed a number of new experimental search strategies adapted to these characteristic final states, also summarised in Table~\ref{tb:strategy}, and implemented the relevant processes in the \Herwigpp event generator in order to study their effectiveness in the presence of QCD radiation, backgrounds and the underlying event.  We used the \texttt{Delphes} detector simulation to assess the effects of $b$ and $\tau$ tagging efficiencies and  detector resolution.
For definiteness we assumed a leptoquark mass of 400 GeV and an integrated $pp$ luminosity of 10 fb$^{-1}$ at 14 TeV.

In the case of decays of leptoquark pairs to $(q\tau)(q\tau)$ where $q=t$ or $b$, the approximate collinearity of the missing neutrinos and jets from the tau decays allows full reconstruction of the leptoquark mass, even when one top decay is semileptonic.
In the former case there is a quartic ambiguity in the resulting mass, although not all of the solutions are real.  After detector resolution smearing, the correct solutions for the momentum fraction $z_2$ may be complex, but we found that using the real parts provides a fair estimate of the mass, with resolution of the order of $\pm 150$ GeV.  For $(b\tau)(b\tau)$ the only ambiguity is combinatoric but the mass resolution is similar.  In both cases the expected background from $q\bar q jj$ is small after cuts and reconstruction.

For decays to $(t\tau)(b\nu)$ or $(t\nu)(b\tau)$, we have proposed an edge reconstruction strategy similar to those developed for supersymmetric models, but using mass variables $M_{\rm min}^{\rm bal}$ and $M_{\rm min}$ that are in principle superior to the classic `stransverse mass' $M_{T2}$.  However, given the limited statistics expected, the difference in performance between these variables was not obvious.  We found cuts to reduce the background from $t\bar t$ to manageable levels, but the edge reconstruction remains challenging without higher statistics.  For $(q\nu)(q\nu)$ the story is similar for edge reconstruction in  $M_{T2}$, the case of $q=t$ being the more difficult owing to the similarity of the distributions of the signal and $t\bar t$ background.  But even in that case a clear excess over background should be visible and would give a rough estimate of the leptoquark mass.

In the event that a discovery is made, one might ask to what extent this provides proof that electroweak symmetry breaking is driven by strongly-coupled, composite dynamics. After all, one can easily imagine weakly-coupled theories with such states, for example, third generation squarks in R-parity-violating supersymmetric models. Ultimately, 
TeV-scale compositeness can only be revealed by experiments probing significantly higher scales; for that, we shall have to wait some time.
In the meantime, the discovery of leptoquarks coupled to third generation fermions and their {\em de facto} consistency with the multitude of existing flavour experiments would imply very strong bounds on the couplings to other fermions. The scenario in which the observed fermions are partially elementary and partially composite provides, as far as we know, the only mechanism in which the required suppression can be automatically achieved. Moreover, it gives a prediction for the size of the other couplings, some of which are not far from current bounds, which may then be targeted in ongoing flavour experiments. Though circumstantial, this would seem to be the best possible evidence for compositeness that one might hope for in the LHC era. 

\section*{Acknowledgements} 
We thank the Herwig++ collaboration and particularly Peter Richardson for help with the Herwig++ implementation. AP would like to thank Andy Pilkington for helpful comments and suggestions.  This work was supported in part by the UK Science and Technology Facilities Council and the Marie Curie Research Training Network ``MCnet'' (contract number MRTN-CT-2006-035606).

\appendix
\section{Feynman rules and diagrams}\label{app:feynman}

\begin{figure}[!htb]
\begin{tabular}{cc}
\begin{minipage}{0.5\hsize}
\begin{center}
  \leavevmode
    \includegraphics[scale=1.00]{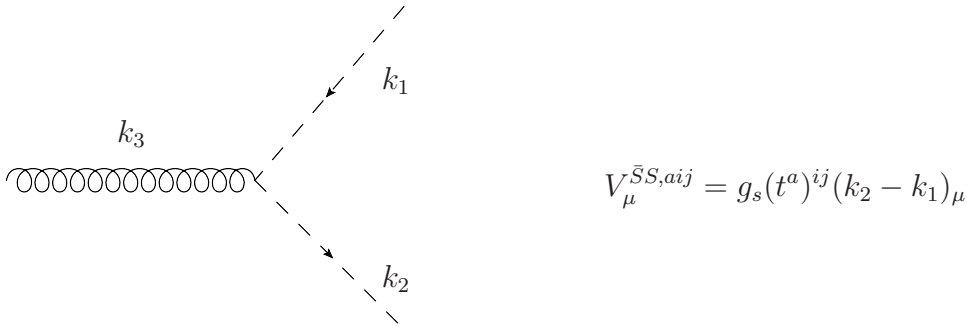}
\put(-110,70){$k_3$}
\put(-10,90){$k_1$}
\put(-10,15){$k_2$}

\end{center}
\end{minipage}
\begin{minipage}{0.5\hsize}
\begin{center}
\begin{equation}
V^{\bar{S}{S},aij}_{\mu} = g_s (t^a)^{ij} ( k_2 - k_1 )_\mu \nonumber
\end{equation}
\end{center}
\end{minipage}
\end{tabular}
\caption{Feynman rule for the vertex scalar leptoquark-scalar anti-leptoquark-gluon. All momenta are incoming and arrows indicate colour flow.}
\label{fig:scalarfeyngss}
\end{figure}

\begin{figure}[!htb]
\begin{tabular}{cc}
\begin{minipage}{0.5\hsize}
\begin{center}
  \leavevmode
    \includegraphics[scale=0.85]{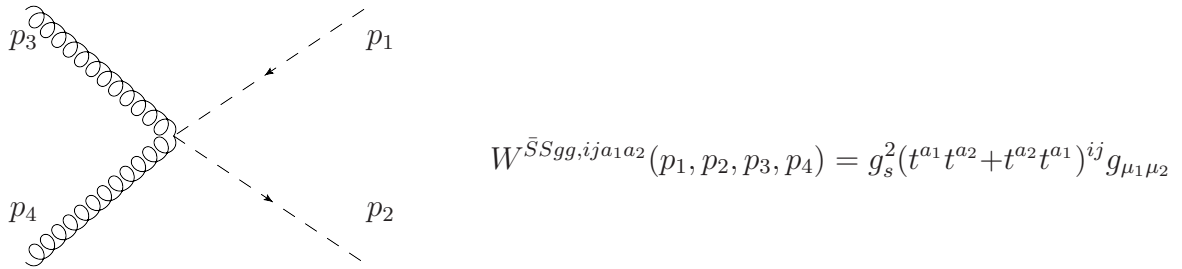}
\put(-135,85){$p_3$}
\put(-135,20){$p_4$}
\put(0,85){$p_1$}
\put(0,20){$p_2$}
\end{center}
\end{minipage}
\begin{minipage}{0.5\hsize}
\begin{center}
\begin{equation}
W^{\bar{S}Sgg,ija_1a_2}(p_1,p_2,p_3,p_4) = g_s^2 ( t^{a_1}t^{a_2} + t^{a_2} t^{a_1})^{ij} g_{\mu _1 \mu_2} \nonumber
\end{equation}
\end{center}
\end{minipage}
\end{tabular}
\caption{Feynman rule for the vertex scalar leptoquark-scalar anti-leptoquark-gluon-gluon. All momenta are incoming.}
\label{fig:scalarfeynggss}
\end{figure}

\begin{figure}[!htb]
 \centering 
    \includegraphics[scale=1.00]{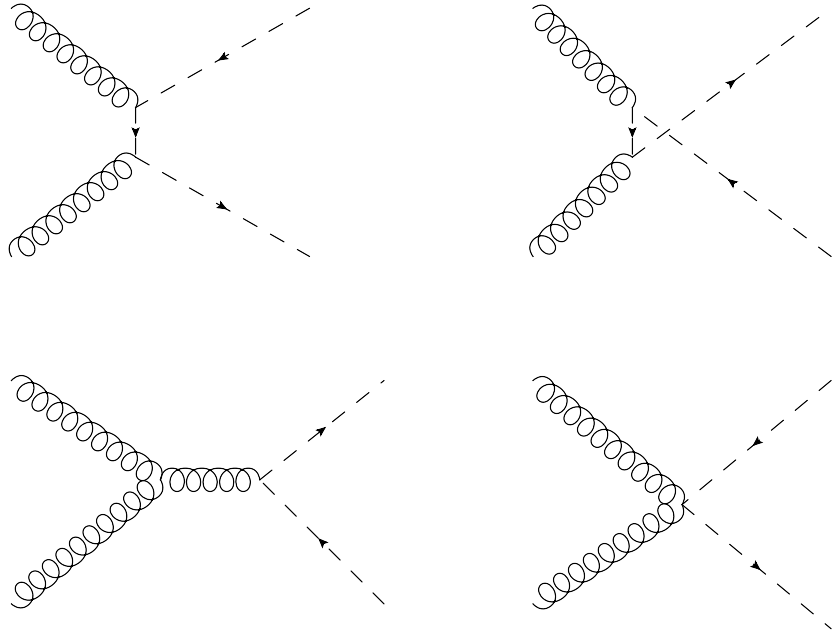}
\caption{Feynman diagrams relevant to scalar leptoquark pair-production with gluon-gluon initial states.}
\label{fig:ggfeynpair}
\end{figure}

\begin{figure}[!htb]
 \centering 
    \includegraphics[scale=1.00]{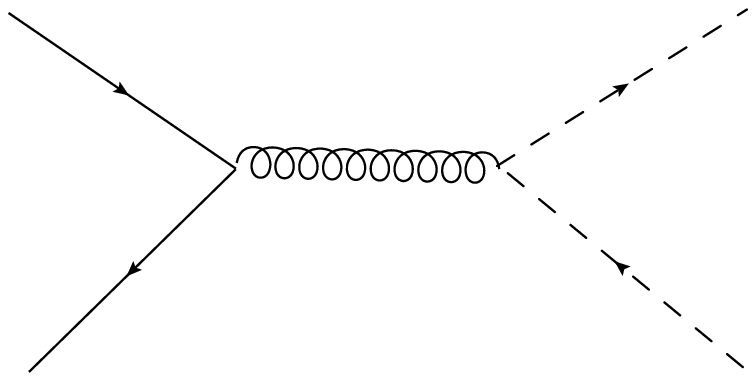}
\caption{Feynman diagram relevant to scalar leptoquark pair-production with guark-antiquark initial states.}
\label{fig:qqfeynpair}
\end{figure}

The Feynman rules~\cite{Blumlein:1996qp} relevant to the leptoquark pair-production diagrams are given in Figure~\ref{fig:scalarfeyngss} and Figure~\ref{fig:scalarfeynggss}. The relevant parton-level Feynman diagrams are shown in Figure~\ref{fig:ggfeynpair} and Figure~\ref{fig:qqfeynpair} for gluon-gluon and quark-antiquark initial states respectively. 


\begin{figure}[!htb]
 \centering 
    \includegraphics[scale=1.00]{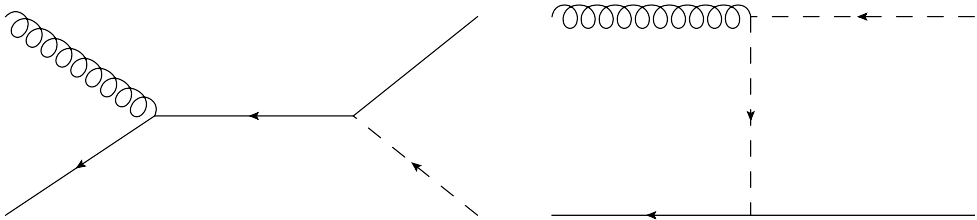}
\caption{Feynman diagrams relevant to scalar leptoquark single production. Solid lines with an arrow indicate quarks, lines without an arrow indicate leptons.}
\label{fig:singlelq}
\end{figure}

Since the couplings to light generations are suppressed in the kind of models we are considering, leptoquark single-production in hadron colliders can proceed only via $b$-quark gluon fusion, as shown in Figure~\ref{fig:singlelq}.  However this is also heavily suppressed due to the low $b$-quark PDF and the small couplings to fermions, and can be neglected.

\section{Cross sections}\label{app:xsections}
Using the given Feynman rules, the cross section for scalar leptoquark pair-production is completely determined, with the only unknown parameter being their mass. The partonic differential cross sections with respect to the leptoquark scattering angle in the partonic centre of mass frame, $\theta$ are given by:
\begin{eqnarray}
\frac{\mathrm{d} \hat{\sigma}^{gg}_{S\bar{S}}} { \mathrm{d} \cos \theta } &=& \frac{ \pi \alpha _s ^2 } { 6 \hat{s} } \beta \left\{ \frac{1}{32} [ 25 + 9 \beta ^2 \cos ^2 \theta - 18 \beta^2 ] \right.\nonumber \\
 &-& \left. \frac{1}{16} \frac{ (25 - 34 \beta ^2 + 9 \beta^4 ) } { 1 - \beta^2 \cos ^2 \theta } + \frac{ ( 1 - \beta ^2 ) ^ 2} { (1 - \beta^2 \cos ^2 \theta )^2} \right\}\;\;, \nonumber \\
\frac{\mathrm{d} \hat{\sigma}^{q\bar{q}}_{S\bar{S}}} { \mathrm{d} \cos \theta } &=& \frac{ \pi \alpha _s ^ 2} { 18 \hat{s} } \beta^3 \sin ^2 \theta \;\;,
\end{eqnarray}
where the `boost factor', $\beta$, is given by $\beta = \sqrt { 1 - 4 M_{LQ}^2 / \hat{s} }$, with $\hat{s}$ the square of the parton-parton centre of mass energy. The integrated partonic cross sections are then given by:
\begin{eqnarray}
 \hat{\sigma}^{gg}_{S\bar{S}} &=& \frac{ \pi \alpha _s ^2 }{96 \hat{s} } \left\{ \beta ( 41 - 31 \beta ^2 ) - (17 - 18 \beta ^2 + \beta^4) \log \left| \frac{1 + \beta} { 1 - \beta } \right| \right\}\;\;, \nonumber \\
\hat{\sigma}^{q\bar{q}}_{S\bar{S}} &=& \frac{ 2 \pi \alpha _s ^ 2 } { 27 \hat{s} } \beta^3 \;\;.
\end{eqnarray}
Note that the production cross sections at leading order in QCD are
exactly equivalent to the supersymmetric top partner pair-production
cross sections. This fact was verified directly using the implemented
supersymmetric model in \Herwigpp. 

We obtain the hadronic cross sections in the usual way, i.e. by folding with the parton density functions and integrating with respect to the momentum fractions $x_{1,2}$:
\begin{eqnarray}
\sigma^{ij}_{S\bar{S}} = \int \mathrm{d} x_1 \mathrm{d} x_2 f_{h_1,i}(x_1,\hat{s}) f_{h_2,j}(x_2,\hat{s}) \hat{\sigma}^{ij}_{S\bar{S}}(\hat{s}) \;\;,
\label{eq:sigmahad}
\end{eqnarray}
where $E_{h_1,h_2}$ is the hadronic centre-of-mass energy, $\hat{s} = x_1 x_2 E_{h_1,h_2}^2$, and $f_{h_a,i}(x_k,\hat{s})$ is the distribution function for parton $i$ in hadron $h_a$, evaluated at scale $\hat{s}$ and momentum fraction $x_k$.

\begin{figure}[!htb]
  \centering 
  \vspace{1.5cm}
    \includegraphics[scale=0.50, angle=90]{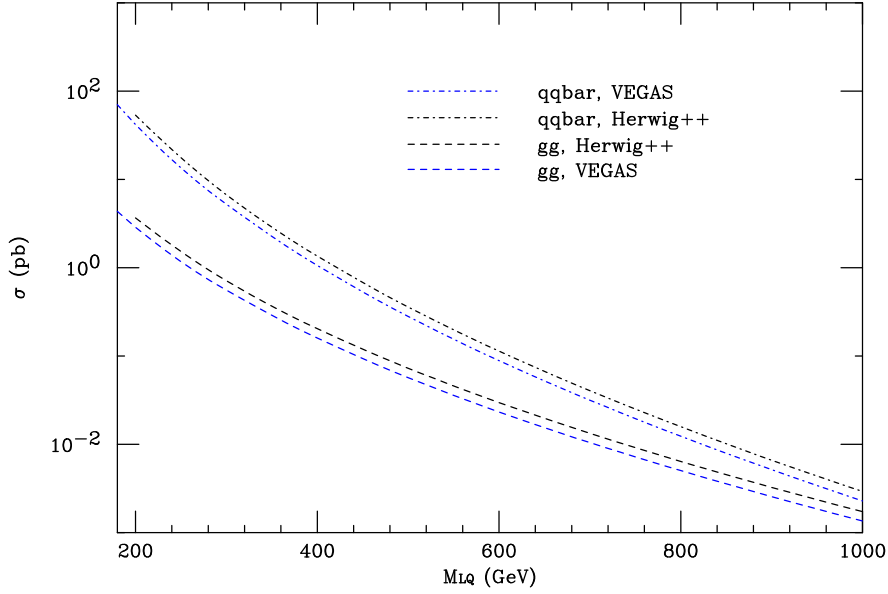}
\caption{The variation of cross section for scalar leptoquark pair-production at the LHC (14 TeV pp centre-of-mass energy) using \Herwigpp and direct integration using the VEGAS algorithm (MRSTMCal parton density functions). }
\label{fig:sigma_slqMnew}
\end{figure}

The vertices appearing in Figure~\ref{fig:scalarfeynggss} and
Figure~\ref{fig:scalarfeyngss} have been implemented in the
\Herwigpp event generator~\cite{Bahr:2008pv,Bahr:2008tf} and the diagrams
contributing to the scalar leptoquark pair-production are reproduced
automatically therein. This yields a calculation  for the cross section which, as a check on the implementation, we compare in Figure~\ref{fig:sigma_slqMnew} with direct integration of Eq.~(\ref{eq:sigmahad}) using the adaptive integration algorithm VEGAS~\cite{Lepage:1980dq}.  For this comparison we used the MRSTMcal (also known as LO**) parton distribution functions~\cite{Sherstnev:2007nd} in both cases.  The slight discrepancy is due to the difference between the internal definition of the
strong coupling constant in \Herwigpp and the one associated with the
MRSTMcal  PDFs.

\section{The effective Lagrangian for decays of derivatively-coupled leptoquarks}\label{app:conj}
The Lagrangian for derivatively-coupled conjugate fields, which appears in Eq.~(\ref{eq:lagd}), also contains terms involving the conjugate fields, such as:
\begin{equation}
\mathcal{L}_{\tilde{S}'_{1/2}} \sim \bar{t}^c_R  \gamma_\mu \tau_Lp^{\mu,q} \tilde{S}'^{(+)}_{1/2}\;\;.
\end{equation}
To manipulate the above expression for the case of on-shell $\tilde{S}'_{1/2}$ decays as we did in Eq.~(\ref{eq:noconjmanip}), we need to show that:
\begin{eqnarray}
\bar{\Psi}^C_{R,L} \slashed{p} = m \bar{\Psi}^C_{L,R}\;\;,
\label{eq:conjmanip1}
\end{eqnarray}
where $\Psi$ is a 4-component spinor and $\Psi_{L,R}^C = (P_{L,R} \Psi)^C$. This can be done by using the following identities~\cite{Dreiner:2008tw}:
\begin{eqnarray}
\bar{\Psi}^C &=& - \Psi ^T C^{-1} \nonumber\\
C^{-1} \gamma_\mu &=& - \gamma_\mu^T C^{-1}\;\;,
\end{eqnarray}
and hence:
\begin{eqnarray}
\bar{\Psi}^C_{R,L} = - \left[P_{R,L} \Psi\right]^T C^{-1} \;\;.
\end{eqnarray}
So the necessary effective Lagrangian for the decay is given by:
\begin{eqnarray}
\mathcal{L}_{eff} \sim m_t \bar{t}^c_L  \tau_L \tilde{S}'^{(+)}_{1/2} \;\;,
\end{eqnarray}
as required.
The full list of effective Lagrangians for the primed leptoquarks, from which the decay modes and couplings in Tables~\ref{tb:lquarksprime} and~\ref{tb:lambdasprime} can be derived, is given by:
\begin{eqnarray}
\mathcal{L}_{S_0'} &=& \left[ \frac{-i }{\sqrt{2} f}  (g'_{0L} m_b+ g'_{0R} m_\tau)\right] \bar{b}_R S_0' \tau_L\nonumber\\
&+& \left[ \frac{-i }{\sqrt{2} f} ( g'_{0L} m_\tau + g'_{0R} m_b )  \right] \bar{b}_L S_0' \tau_R \nonumber \\
&+& \left[  \frac{-i }{\sqrt{2} f} (g'_{0L} m_t)\right]  \bar{t}_R S_0' \nu_{\tau,L}\;\;,\\
\mathcal{L}_{\tilde{S}_0'} &=&  \left[ \frac{-i }{\sqrt{2} f} (\tilde{g}'_{0R} m_t \bar{t}_L \tau_R + \tilde{g}'_{0R} m_\tau \bar{t}_R \tau_L) \tilde{S}'_0 \right]\;\;,\\
\mathcal{L}_{S_1'} &=&  \left[ \frac{-i }{\sqrt{2} f} \sqrt{2} g'_{1L} ( m_t \bar{t}_R \tau_L + m_\tau \bar{t}_L \tau_R ) S'^{(+)}_1 \right. \nonumber \\
&+& \frac{-i }{\sqrt{2} f} \sqrt{2} g'_{1L} m_b \bar{b}_R \nu_L S'^{(-)}_1 \nonumber \\
&+& \left. \frac{-i }{\sqrt{2} f} ( g'_{1L} m_t \bar{t}_R \nu _L - g'_{1L} m_b \bar{b}_R \tau_L - g'_{1L} m_\tau \bar{b}_L \tau_R) S'^{(0)}_1 \right] \;\;,\\
\mathcal{L}_{S_{1/2}'} &=& \left[ \frac{-i }{\sqrt{2} f} (h'_{1L} m_b \bar{b}^c_L \nu_L + h'_{1R} m_t \bar{t}_R^c \tau_R + h'_{1R} m_\tau \bar{t}^c_L \tau_L) S'^{(-)}_{1/2} \right. \nonumber\\
&+& \left. \frac{-i }{\sqrt{2} f} ( h'_{1L} m_b + h'_{1R} m_\tau ) \bar{b}^c_L \tau_L S'^{(+)}_{1/2} + ( h'_{1L} m_\tau + h'_{1R} m_b ) \bar{b}^c_R \tau_R  S'^{(+)}_{1/2} \right] \;\;,\\
\mathcal{L}_{\tilde{S}_{1/2}'} &=& \left[ \frac{-i }{\sqrt{2} f} h'_{2L} m_t \bar{t}_L^c \nu_L \tilde{S}'^{(-)}_{1/2} \right. \nonumber \\
&+& \left. (h'_{2L} m_t \bar{t}^c_L \tau_L + h'_{2L} m_\tau \bar{t}^c_R \tau_R ) \tilde{S}'^{(+)}_{1/2} \right]\;\;,
\end{eqnarray}
where we have defined: $S'^{(\pm)}_1 \equiv (S'^{(1)}_1 \mp
iS'^{(2)}_1)/\sqrt{2}$ (and equivalent definitions for $\tilde{S}'^{(\pm)}_{1/2}$) and $S'^{(0)}_1 \equiv S'^{(3)}_1$. We have also used the
fact that the doublet leptoquarks may be written as a vector $S'_{1/2}
=  ( S'^{(-)}_{1/2},  S'^{(+)}_{1/2} )$. We have set the quark and
lepton couplings to equal, $g^q = g^\ell$,\footnote{The implementation in the upcoming \Herwigpp version 2.5.0 also includes this simplification.} however these can be
reinstated trivially by replacing $g \rightarrow g^q$ where a quark
mass term appears and $g \rightarrow g^\ell$ where a lepton mass term appears. 

Note that terms appearing in this Lagrangian are no longer $SU(2)_L \times U_Y$ gauge-invariant. This is consistent since these terms would appear in the Lagrangian after electroweak symmetry breaking and vanish as the fermion masses tend to zero. The Lagrangian is, of course, $U(1)_{em}$ gauge-invariant.

\section{\boldmath $(t\tau)(t\tau)$ reconstruction method}\label{app:ttauttau} 
In terms of the momentum ratios $z_i$ defined in Eq.~(\ref{eq:zis}), the conditions for balancing the total missing transverse momentum can be written as
\beqn
z_1 &=& (\ptmx - ( z_2 -1) p_{j_2}^x - p_{\nu_l}^x) / p_{j_1}^x + 1, \label{eq:z1}\\
p_{j_1}^y p_{\nu_l}^x - p_{j_1}^x p_{\nu_l}^y  &=&
\ptmx p_{j_1}^y - \ptmy p_{j_1}^x + (z_2 -1) (p_{j_1}^x p_{j_2}^y - p_{j_1}^y p_{j_2}^x) .
\label{eq:lin1}
\eeqn
The mass-shell conditions, except for $p_{\nu_l}^2=0$, can be written as
\beqn
m_W^2 &=& (p_l + p_{\nu_l})^2 = 2 p_l \cdot p_{\nu_l}\;\;,  \label{eq:lin2} \\
m_t^2 &=& (p_b + p_l + p_{\nu_l})^2 = m_W^2 + m_b^2 + 2 p_b \cdot p_{l} + 2 p_b \cdot p_{\nu_l} \label{eq:lin3}\;\;, \\
m^2_{S_0} &=& (p_t + p_{\tau_1})^2 = \tilde m_t^2 + 2 z_1 p_t \cdot p_{j_1} \label{eq:ms1} \;\;,\\
m^2_{S_0} &=& (p_b + p_l + p_{\nu_l} + p_{\tau_2})^2 = m_t^2 + 2 z_2 (p_b + p_l) \cdot p_{j_2}
+ 2 z_2 p_{j_2} \cdot p_{\nu_l} \label{eq:ms2} \;\;,
\eeqn
where $\tilde{m}_t$ is the reconstructed mass of the hadronic top and $m_t$ is the assumed mass of the semi-leptonic top.
By eliminating $z_1$ and $m_{S_0}$ from eqs.\,(\ref{eq:z1}), (\ref{eq:ms1}) and (\ref{eq:ms2}), 
one obtains
\beq
z_2 p_{j_2}\cdot p_{\nu_l}+
\frac{p_t \cdot p_{j_1}}{p_{j_1}^x} p_{\nu_l}^x  
= t_3 + u_3 z_2,
\label{eq:lin4}
\eeq
where
\beqn
t_3 &=& \frac{\tilde m_t^2 - m_t^2}{2} + 
\frac{\ptmx + p_{j_1}^x + p_{j_2}^x}{p^x_{j_1}}
p_t\cdot p_{j_1}\;\;,  \\
u_3 &=& -(p_b+p_l)\cdot p_{j_2} - \frac{p^x_{j_2}}{p_{j_1}^x} p_t \cdot p_{j_1}\;\;.
\eeqn
Using a vector ${\bf p}_{\nu_l} = (E_{\nu_l}~p_{\nu_l}^x~p_{\nu_l}^y~p_{\nu_l}^z)$,
eqs.\,(\ref{eq:lin1}), (\ref{eq:lin2}), (\ref{eq:lin3}) and (\ref{eq:lin4}) can be recast as
\beq
{\bf A} {\bf P}_{\nu_l} = {\bf S}
\label{eq:aps}
\eeq
where
\beq
{\bf A} = 
\begin{pmatrix}
E_l & -p_l^x & -p_l^y & - p_l^z  \\
E_b & -p_b^x & -p_b^y & -p_b^z \\
z_2 E_{j_2} & -z_2 p_{j_2}^x + (p_t \cdot p_{j_1})/p_{j_1}^x & -z_2 p_{j_1}^y & -z_2 p_{j_2}^z \\
0 & p_{j_1}^y & -p_{j_1}^x & 0
\end{pmatrix},
\eeq
and
\beq
{\bf S} =
\begin{pmatrix}
\frac{m_W^2}{2}, & \frac{m_t^2 - m_b^2-m_W^2}{2} - p_b \cdot p_l, & t_3+u_3 z_2, & t_4 + u_4 z_2
\end{pmatrix}.
\eeq
$t_4$ and $u_4$ are defined as
\beqn
t_4 &=& (\ptmx + p_{j_2}^x) p_{j_1}^y - (\ptmy + p_{j_2}^y) p_{j_1}^x, \\
u_4 &=& p_{j_1}^x p_{j_2}^y - p_{j_1}^y p_{j_2}^x .
\eeqn
From Eq.~(\ref{eq:aps}), one can determine ${\bf p}_{\nu_l}$
as a function of $z_2$.
Finally, one can determine $z_2$ from the mass-shell condition 
\beq
{\bf p}_{\nu_l}^2 = ( {\bf A}^{-1} {\bf S})^2 = 0 .
\eeq
This provides a quartic equation for $z_2$, and
one can find up to four real solutions in the physical range $z_2 \ge 1$. 
One can then obtain $m_{S_0}$ by substituting $z_2$ into
Eq.~(\ref{eq:ms2}). 

\section{\boldmath $(q'\tau)(q \nu)$ reconstruction method}\label{app:mminbal}
Given $w$ in Eq.~(\ref{eq:mbnu}), one can minimise $m_{b\nu}(w,p_{\nu}^z)$ in terms of $p_{\nu}^z$. 
The result is: 
\beqn
[m_{b\nu}^{\rm min}(w)]^2 &=& m^2_{b\nu}(w, \tilde p_{\nu}^z)
\nonumber \\ &=&
2|{\bf p}_b| |{\bf p}_{\rm miss}-w {\bf p}_j| - 2 {\bf p}_b \cdot ({\bf p}_{\rm miss} - w{\bf p}_j)
\nonumber \\ &=&
[m^{b\nu}_{T}(w)]^2,
\label{eq:mbnut}
\eeqn
where
\beq
\tilde p_{\nu}^z \equiv \frac{|{\bf p}_{\rm miss} - w {\bf p}_j|}{|{\bf p}_b|} p_b^z\,
\eeq 
and $m^{b\nu}_{T}(w)$ is the transverse mass of the $b \nu$ system.
This allows us to calculate $M_{\rm min}$ by one parameter minimisation
\beq
M_{\rm min} = \min_w [ \max\{ m_{t\tau}(w), m^{b\nu}_T(w) \}]. 
\eeq
Since $m_{t\tau}(w)$ is a monotonically increasing function of $w$,
if $m_{t\tau}(0) \ge  m^{b\nu}_T(0)$,
\beq
M_{\rm min} = m_{t\tau}(0)\,.
\label{eq:mmin1}
\eeq
Furthermore, since there exists a value $\hat p_{\nu}^z$ which fulfils $m_{t\tau}(0) = m_{b\nu}(0,\hat p_{\nu}^z)$,
one finds
\beq
M_{\rm min}^{\rm bal} = m_{t\tau}(0)
\label{eq:mminb1}
\eeq
If $m^{b\nu}_T(0) > m_{t\tau}(0)$, we have to search for other values of $w$.
For the true $w$ and $p_{\nu}^z$, say $w^*$ and $p_{\nu}^{z*}$, we have 
\beq
m^{b\nu}_T(w^*) < m_{b\nu}(w^*,p_{\nu}^{z*}) = m_{t\tau}(w^*)\,.
\eeq
This assures existence of $\hat w$ which satisfies the relation $m^{b\nu}_T(\hat w)=m_{t\tau}(\hat w)$.
By scanning $w$ from 0 to $\hat w$, one finds
\beq
M_{\rm min}^{\rm bal} = m^{b\nu}_T(\hat w)\,,
\label{eq:mmin2}
\eeq
and
\beq
M_{\rm min} = \min_{w\in[0-\hat w]} [m^{b\nu}_T(w)]\,.
\label{eq:mminb2}
\eeq
Hence we have:
\beq
M_{\rm min}^{\rm bal} \ge M_{\rm min}.
\eeq
\bibliography{leptoquarks.bib}
\bibliographystyle{utphys}

\end{document}